\documentclass[12pt]{iopart}
\usepackage{iopams} 
\usepackage{graphics}
\usepackage{pennames}
\usepackage{amssymb}
\usepackage{setstack}
\def\La{\Lambda}
\def\Al{\overline\Lambda}
\def\XI{\Xi^{-}}

\def\mt{m_{\tt T}}
\def\Bt{\langle\beta_\perp\rangle}
\def\pt{p_{\tt T}}
\def\diffD{{\rm d}}
\newcommand{\decayarrow}{\makebox[0mm][l]{\rule{0.33em}{0mm}\rule[0.55ex]{0.044em}{1.55ex}}\rightarrow}
\begin{document}
\title[Study of the $m_{\tt T}$\ spectra of strange particles in 
       Pb--Pb at 40 $A$\ GeV/$c$]
%{Transverse dynamics of strange hadrons in  
% lead--lead collisions at 40 $A$\ GeV/$c$\ beam momenta}  
{Transverse dynamics of Pb--Pb collisions at 40~$A$~GeV/$c$\ viewed by strange hadrons}
\author{  
F~Antinori$^{1}$,
P~Bacon$^{2}$,
A~Badal{\`a}$^{3}$,
R~Barbera$^{3}$,
A~Belogianni$^{4}$,
I~J~Bloodworth$^{2}$,
M~Bombara$^{5}$,
G~E~Bruno$^{6}$,
%\footnote[1]{To
%whom correspondence should be addressed (giuseppe.bruno@ba.infn.it)},
S~A~Bull$^{2}$,
R~Caliandro$^{6}$,
M~Campbell$^{7}$,
W~Carena$^{7}$,
N~Carrer$^{7}$,
R~F~Clarke$^{2}$,
A~Dainese$^{1}$\footnote[2]{Present address: Laboratori Nazionali di Legnaro, Legnaro, Italy},
D~Di~Bari$^{6}$,
S~Di~Liberto$^{8}$,
R~Divi\`a$^{7}$,
D~Elia$^{6}$,
D~Evans$^{2}$,
G~A~Feofilov$^{9}$,
R~A~Fini$^{6}$,
P~Ganoti$^{4}$,
B~Ghidini$^{6}$,
G~Grella$^{10}$,
H~Helstrup$^{11}$,
K~F~Hetland$^{11}$,
A~K~Holme$^{12}$,
A~Jacholkowski$^{3}$,
G~T~Jones$^{2}$,
P~Jovanovic$^{2}$,
A~Jusko$^{2}$,
R~Kamermans$^{13}$,
J~B~Kinson$^{2}$,
K~Knudson$^{7}$,
V~Kondratiev$^{9}$,
I~Kr\'alik$^{5}$,
A~Krav\v c\'akov\'a$^{14}$,
P~Kuijer$^{13}$,
V~Lenti$^{6}$,
R~Lietava$^{2}$,
G~L\o vh\o iden$^{12}$,
V~Manzari$^{6}$,
M~A~Mazzoni$^{8}$,
F~Meddi$^{8}$,
A~Michalon$^{15}$,
M~Morando$^{1}$,
P~I~Norman$^{2}$,
A~Palmeri$^{3}$,
G~S~Pappalardo$^{3}$,
B~Pastir\v c\'ak$^{5}$,
R~J~Platt$^{2}$,
E~Quercigh$^{1}$,
F~Riggi$^{3}$,
D~R\"ohrich$^{16}$,
G~Romano$^{10}$,
R~Romita$^{6}$,
K~\v{S}afa\v{r}\'{\i}k$^{7}$,
L~\v S\'andor$^{5}$,
E~Schillings$^{13}$,
G~Segato$^{1}$,
M~Sen\'e$^{17}$,
R~Sen\'e$^{17}$,
W~Snoeys$^{7}$,
F~Soramel$^{1}$\footnote[1]{Permanent
address: University of Udine, Udine, Italy},
M~Spyropoulou-Stassinaki$^{4}$,
P~Staroba$^{18}$,
R~Turrisi$^{1}$,
T~S~Tveter$^{12}$,
J~Urb\'{a}n$^{14}$,
P~van~de~Ven$^{13}$,
P~Vande~Vyvre$^{7}$,
A~Vascotto$^{7}$,
T~Vik$^{12}$,
O~Villalobos~Baillie$^{2}$,
L~Vinogradov$^{9}$,
T~Virgili$^{10}$,
M~F~Votruba$^{2}$,
J~Vrl\'{a}kov\'{a}$^{14}$\ and
P~Z\'{a}vada$^{18}$
}
\address{
$^{1}$ University of Padua and INFN, Padua, Italy\\
$^{2}$ University of Birmingham, Birmingham, UK\\
$^{3}$ University of Catania and INFN, Catania, Italy\\
$^{4}$ Physics Department, University of Athens, Athens, Greece\\
$^{5}$ Institute of Experimental Physics, Slovak Academy of Science,
              Ko\v{s}ice, Slovakia\\
$^{6}$ Dipartimento IA di Fisica dell'Universit{\`a}
       e del Politecnico di Bari and INFN, Bari, Italy \\
$^{7}$ CERN, European Laboratory for Particle Physics, Geneva, Switzerland\\
$^{8}$ University ``La Sapienza'' and INFN, Rome, Italy\\
$^{9}$ State University of St. Petersburg, St. Petersburg, Russia\\
$^{10}$ Dipartimento di Scienze Fisiche ``E.R. Caianiello''
       dell'Universit{\`a} and INFN, Salerno, Italy\\
$^{11}$ H{\o}gskolen i Bergen, Bergen, Norway\\
$^{12}$ Fysisk Institutt, Universitetet i Oslo, Oslo, Norway\\
$^{13}$ Utrecht University and NIKHEF, Utrecht, The Netherlands\\
$^{14}$ P.J. \v{S}af\'{a}rik University, Ko\v{s}ice, Slovakia\\
$^{15}$ IReS/ULP, Strasbourg, France\\
$^{16}$ Fysisk Institutt, Universitetet i Bergen, Bergen, Norway\\
$^{17}$ Coll\`ege de France, Paris, France\\
$^{18}$ Institute of Physics, Prague, Czech Republic 
}
\ead{Giuseppe.Bruno@ba.infn.it}
\begin{abstract}
The transverse mass spectra of \PKzS, \PgL, $\Xi$\ and  $\Omega$\ particles 
produced  in Pb--Pb collisions at 40 $A$\ GeV/$c$\   
have been studied for a sample of events corresponding to the most 
central 53\% of the inelastic Pb--Pb cross-section.    
We analyze the distributions in the framework of 
%the `blast-wave'' 
a parameterized model inspired by hydrodynamics.  
The dependence of the freeze-out parameters on  
particle species and event centrality is discussed and comparisons with 
results at higher energy are shown.   
\end{abstract}
%\submitto{\JPG}
%Version.1  20/12/2005
%\maketitle
%
\section{Introduction} 
Ultra-relativistic collisions between 
heavy ions are used to study the properties of nuclear matter 
at high energy density. In particular, lattice Quantum Chromo-Dynamics  
(QCD) calculations predict a state of matter of deconfined quarks and gluons (quark-gluon plasma, QGP) 
at an energy density exceeding $\sim$\ %1 
0.6  
GeV/fm$^3$~\cite{lattice}.   
For recent reviews of experimental results and theoretical developments 
see references~\cite{QM04-QM05}.  

Strange particle production has proved to be a powerful tool to study 
the system formed in  
heavy ion collisions~\cite{rafSQM03}. If a QGP state is formed, an  
increased production of ${\rm s}$\ and ${\rm \bar{s}}$\ 
quarks with respect to normal hadronic interactions is expected~\cite{rafelski}.  
The abundance of ${\rm s}$\ and ${\rm \bar{s}}$\ quarks %is expected to 
would 
equilibrate in a few fm/$c$, comparable with the plasma lifetime. The result,  
after statistical hadronization, would be an enhancement of the strange and multi-strange 
particle abundance in nucleus--nucleus interactions with respect to %a superposition of individual 
nucleon-nucleon interactions; 
the enhancement was also predicted to increase with the 
%content of valence strange quarks. 
number of valence strange quarks~\cite{rafelski}.  
%NA57 and its predecessor WA97 have reported~\cite{enh160,WA97enh} an enhanced production of strange particles  
%in Pb--Pb collisions at 158 $A$\ GeV/$c$\ with the predicted hierarchy expected in a QGP scenario. 
We have reported recently~\cite{enh160} results confirming the WA97 finding~\cite{WA97enh}  
that strange particle production in Pb--Pb collisions  at 158 $A$~GeV/$c$\ is enhanced, 
with the predicted hierarchy as expected in a QGP scenario.  
The effect amounts to about a factor 20 enhancement for the triply-strange $\Omega$.  
Preliminary results at 40 $A$\ GeV/$c$~\cite{NA57QM04}   
indicate a similar pattern of the strangeness enhancement to that observed at 
higher energy. Enhancement of strange meson and baryon production in Pb--Pb collisions 
has also been reported by the NA49 Collaboration~\cite{NA49pions,NA49Lambda,NA49str}.  

%If the dense matter formed in the nuclear collision thermalizes locally on a time 
%scale much shorter than any macroscopic dynamical scale, its evolution can be  
%described by ideal fluid dynamics.  
%In the last stage of the collision, the dynamics of the hadronic fireball 
%can be studied from the details of the kinematical distributions of the 
%emerging particles.  
Ratios of particle abundances are rather well described by thermal models~\cite{ThermalMod};
this is interpreted as strong evidence of local thermalization of the nuclear
matter formed in the collision on a time scale shorter than %any macroscopic dynamical scale.
the lifetime of the fireball.  
Local thermal equilibrium is a prerequisite to allow the description of the evolution 
of the system using ideal fluid dynamics,   
provided that the mean free path of the fireball constituents is short enough.  
%In the last stage of the collision, the dynamics of the hadronic fireball
%can be studied from the details of the kinematical distributions of the
%emerging particles. 
Hydro-dynamics or parameterized models inspired by hydro-dynamics have
shown to give a successful description of a number of observables,  
i.e. transverse momentum ($p_{\tt T}$) and rapidity ($y$) distributions,
direct and elliptic flow, and two-particle correlation functions 
(for recent reviews see,  e.g.,  references~\cite{ReviewHydro}).

Collective dynamics in the transverse direction is of major interest
since it can only arise by the buildup of a pressure gradient 
%in that direction 
which, in turn, would be strongly suggestive of thermal  
equilibration %of the nuclear matter.  
of  the system formed in the collision.  

The shapes of the $p_{\tt T}$\ spectra are expected to be determined by 
%the superposition of two effects: 
an interplay between two effects:  
the thermal motion of the
particles in the fireball and a pressure-driven radial flow, induced by the
fireball expansion.
To disentangle the two contributions  we rely on the blast-wave  
model~\cite{BlastRef},
which assumes cylindrical symmetry for an expanding fireball in local
thermal equilibrium, under different hypotheses on the transverse
flow profile.
The analysis of transverse expansion in Pb--Pb at 158 $A$\ GeV/$c$\ was 
presented in reference~\cite{BlastPaper}.  
%The transverse mass ($m_{\tt T}=\sqrt{p_{\tt T}^2+m^2}$)  
%spectra are expected to be sensitive to the details  
%of the production  dynamics~\cite{BlastRef,BlastRef2}.  
In this paper we discuss the study of the  
transverse mass ($m_{\tt T}=\sqrt{p_{\tt T}^2+m^2}$) 
spectra for  \PgL, \PgXm, \PgOm\ hyperons,  
their anti-particles and \PKzS\ 
measured in Pb--Pb collisions at 40 $A$\ GeV/$c$.  
\section{The NA57 set-up}
\begin{figure}[hbt]
\centering
\resizebox{0.88\textwidth}{!}{%
\includegraphics{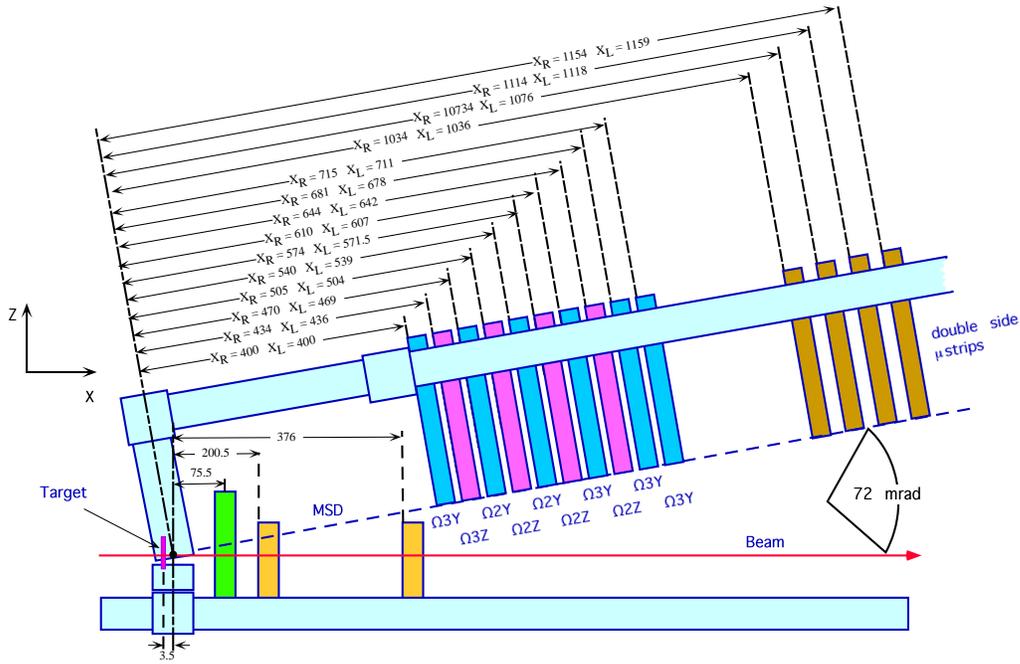}}
\caption{Sketch (not to scale) of the NA57 apparatus placed inside 
the 1.4 T field of the GOLIATH
 magnet in the 40 $A$\ GeV/$c$\ set-up. All distances are given in mm.}
\label{fig:setup}
\end{figure}

The \PgL, \PgXm, \PgOm\ hyperons, their anti-particles and the \PKzS\ mesons
are identified by reconstructing their weak  
decays into final states containing only charged particles:
%e.g.: $\Xi^-$ $\rightarrow$ $\Lambda\pi^-$, with
%$\Lambda$ $\rightarrow$ $\pi^-{p}$.
\begin{equation}
\label{eq:decay}
\begin{array}{lllllll}
    \PKzS &\rightarrow & \pi^+ + \pi^-  & \hspace{10mm}&
    \La  &\rightarrow & p + \pi^-                             \\ \\
    \XI &\rightarrow & \La + \pi^- &  \hspace{10mm}&  \Omega^- &\rightarrow &
   \La + K^-  \\
            &            &
   \decayarrow  p + \pi^-  &  &  &  & \decayarrow  p + \pi^-   \\
\end{array}
\end{equation}
and the corresponding charge conjugates for anti-hyperons.
The charged tracks emerging from strange particle decays 
were reconstructed  
in a telescope made %from 
of  
an array of silicon detector planes of 5x5 cm$^2$\ cross-section
placed in an approximately uniform magnetic field of 1.4 Tesla perpendicular to the 
beam line; the bulk of the detectors was closely packed in an approximately 30 cm long  
compact part used for pattern recognition.  
Descriptions of the NA57 apparatus %in the 158 $A$\ GeV/$c$\ beam momentum set-up 
can be found in references~\cite{enh160,BlastPaper,MANZ}. There, however, emphasis is 
placed on the 158 $A$\ GeV/$c$\ beam momentum set-up. 
The experimental set-up for Pb-Pb collisions at 40 $A$\ GeV/$c$\ 
is conceptually similar: the only differences are in the telescope inclination and 
detector positions which have been chosen in order to optimize at the lower energy the acceptance 
and reconstruction efficiency of strange particles produced at mid-rapidity.  
A sketch of the silicon telescope as installed on the NA57 optical bench  
is shown in figure~\ref{fig:setup}. 
% B. Ghidini suggests to skip this description.
%The compact part of the telescope was made using ten hybrid silicon pixel detectors, with 
%the long pixel direction 
%alternately oriented along the magnetic field and along the bend direction 
%($z$\ and $y$, respectively, in figure~\ref{fig:setup}). Two types of silicon pixel 
%detectors were used: five Omega2 type planes~\cite{Omega2}    
%(indicated in figure~\ref{fig:setup} with symbols ${\rm \Omega2Y}$\ and 
%${\rm \Omega2Z}$\ for $y$\ and $z$\ planes, respectively) with a pixel size of 
%$75 \times 500 \, \mu {\rm m}^2$, and five Omega3 type planes~\cite{Omega3}  
%(${\rm \Omega3Y}$\ and ${\rm \Omega3Z}$, in figure~\ref{fig:setup}), with a pixel 
%size of $50 \times 500 \, \mu {\rm m}^2$. 
%To improve the momentum resolution of high momentum tracks a
%lever arm detector (an array of four double-sided silicon micro-strip detectors)  
%was placed downstream of the tracking telescope.  
%The whole silicon telescope was placed in the GOLIATH magnet, 
%above the beam line, inclined and aligned 
%with the lower edge of the detectors laying on a line pointing back to the target.
An inclination angle of $72$\ mrad with respect to the beam line 
and a distance of the first pixel plane from the target equal to $40$\ cm were     
set in order to accept particles produced in about half a unit of rapidity 
around central rapidity at medium transverse momentum.  

An array of  scintillation counters (Petals), placed  7.9 cm downstream of the target, 
provided a fast signal to trigger on the centrality of the collisions. 
The Petals covered the pseudo-rapidity region $0.8<\eta<1.8$\   
% my computation is: 0.829 -- 1.775
%, thus not shadowing the  the silicon telescope, 
and their thresholds could be set so as to accept events with track multiplicities above an 
adjustable limit. This was tuned so that the triggered event sample 
corresponds to approximately 
the most central %60\% 
56\% of the Pb--Pb inelastic cross-section. 
The thickness of the Pb target was 1\% of an interaction length.  

The centrality of the Pb-Pb collisions is determined (off-line) by analyzing the
charged particle multiplicity measured by two stations of micro-strip silicon 
detectors (MSD) which sample the pseudo-rapidity intervals $1.9<\eta<3$\ and $2.4<\eta<3.6$.  
%Each station consists of three arms, each of them composed of 200 strips of pitches 
%ranging from 100 μm to 400 μm.  The stations are positioned at
%20.4 cm and 38.0 cm from the target.

\section{Data sample and analysis}
The results presented in this paper are based on the analysis of 
the full data sample collected in Pb--Pb collisions at 40 $A$\ GeV/$c$, 
consisting of 240 M  events. The selected sample 
%of events  % original 
used for the analysis % B. Ghidini
corresponds to the most central 53\% of the inelastic Pb--Pb cross-section.  
%The triggered fraction of the Pb--Pb inelastic cross-section is about 60\%. 
%
%All data are corrected for geometrical acceptance and for detector and 
%reconstruction inefficiencies on an event-by-event  
%basis, with the procedure described in reference~\cite{QM02Manzari}. 
%
The data sample has been divided into five centrality classes (0,1,2,3 and 4,  
class 4 being the most central) according to the value of the charged particle 
multiplicity %around central rapidity measured by a Silicon Microstrip Multiplicity Detector.  
measured by the MSD. 
The procedure for the measurement of the multiplicity distribution and 
the determination of the collision centrality for each class 
is described in reference~\cite{Multiplicity}.  
The fractions of the inelastic cross-section for the five classes, calculated  
assuming %a total 
an  inelastic Pb--Pb  
cross-section of 7.26 barn, are the same as those defined 
at higher beam momentum (158 $A$\ GeV/$c$) and they are given in table~\ref{tab:centrality}.  
\begin{table}[h]
\caption{Centrality ranges for the five classes.
\label{tab:centrality}}
\begin{center}
\begin{tabular}{llllll}
\hline
 Class &   $0$   &   $1$   &   $2$   &  $3$   &   $4$ \\ \hline
 $\sigma/\sigma_{inel}$\ \; (\%)   & 40 to 53 & 23 to 40 & 11 to 23& 4.5 to 11 & 0 to 4.5 \\
 \hline
\end{tabular}
\end{center}
\end{table}
\noindent
%in table~\ref{tab:participants}.  
%\begin{table}[h]
%\caption{Average number of $N_{wound}$\ in  the five  classes defined  
%in Pb--Pb interactions. 
%\label{tab:participants}}
%\begin{center}
%\begin{tabular}{|r|c|c|c|c|c|}
%\hline
%                  &    $0$   &   $1$     &    $2$    &   $3$     &   $4$ \\ \hline
% 40 $A$\ GeV/$c$  & $57\pm5$ & $119\pm5$ & $208\pm4$ & $292\pm1$ & $346\pm1$ \\
%158 $A$\ GeV/$c$  & $62\pm4$ & $121\pm4$ & $209\pm3$ & $290\pm2$ & $349\pm1$ \\
%%                       $0$   &   $1$     &    $2$    &   $3$     &   $4$ \\ \hline
%%$58\pm4$ & $117\pm4$ & $204\pm3$ & $287\pm2$ & $349\pm1$ \\
%\hline
%\end{tabular}
%\end{center}
%\end{table}
%\noindent
%\vspace{-1.0cm}

\begin{figure}[hbt]
\centering
\resizebox{0.65\textwidth}{!}{%
\includegraphics{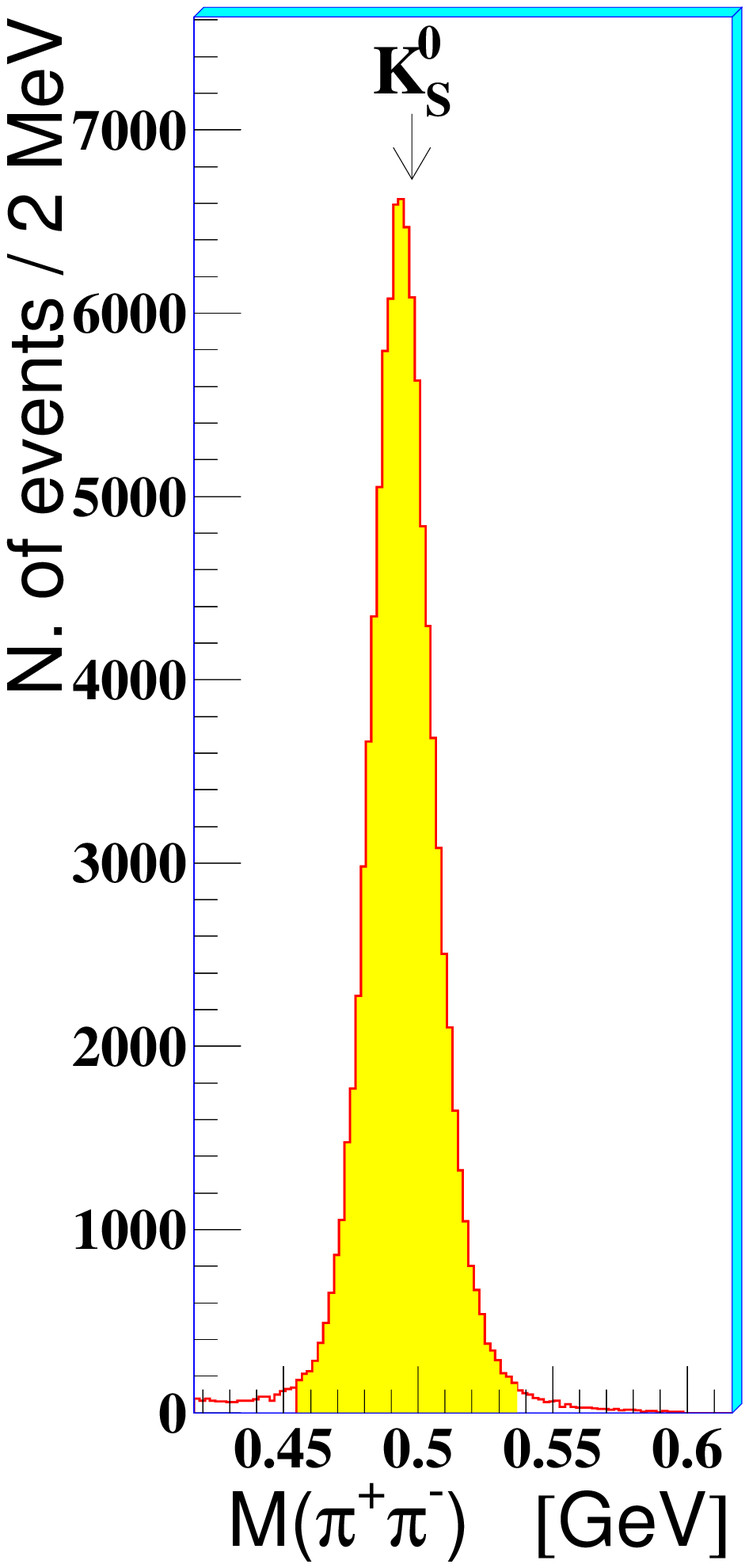}
\includegraphics{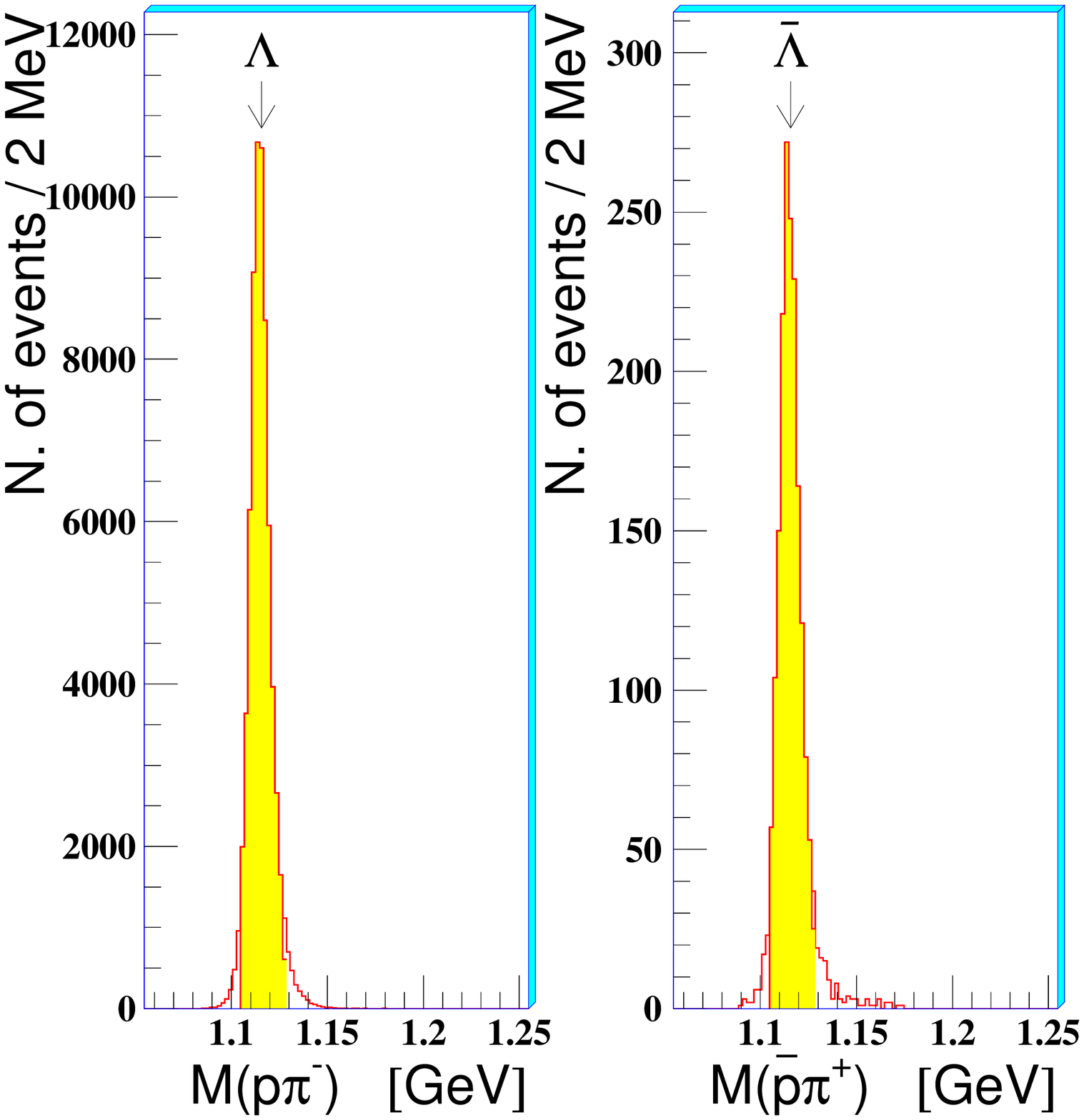}}\\
\resizebox{0.84\textwidth}{!}{%
\includegraphics{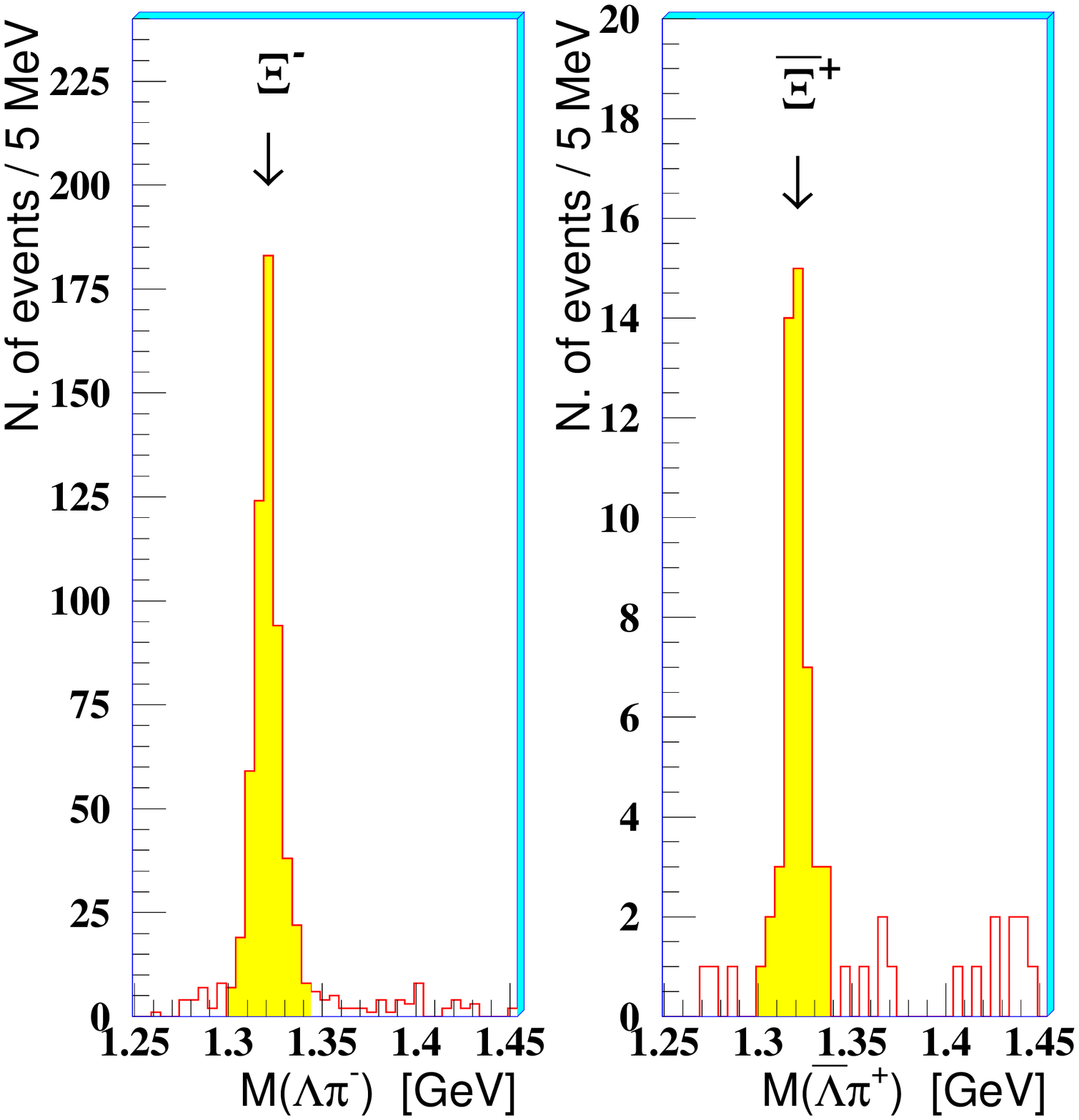}
\includegraphics{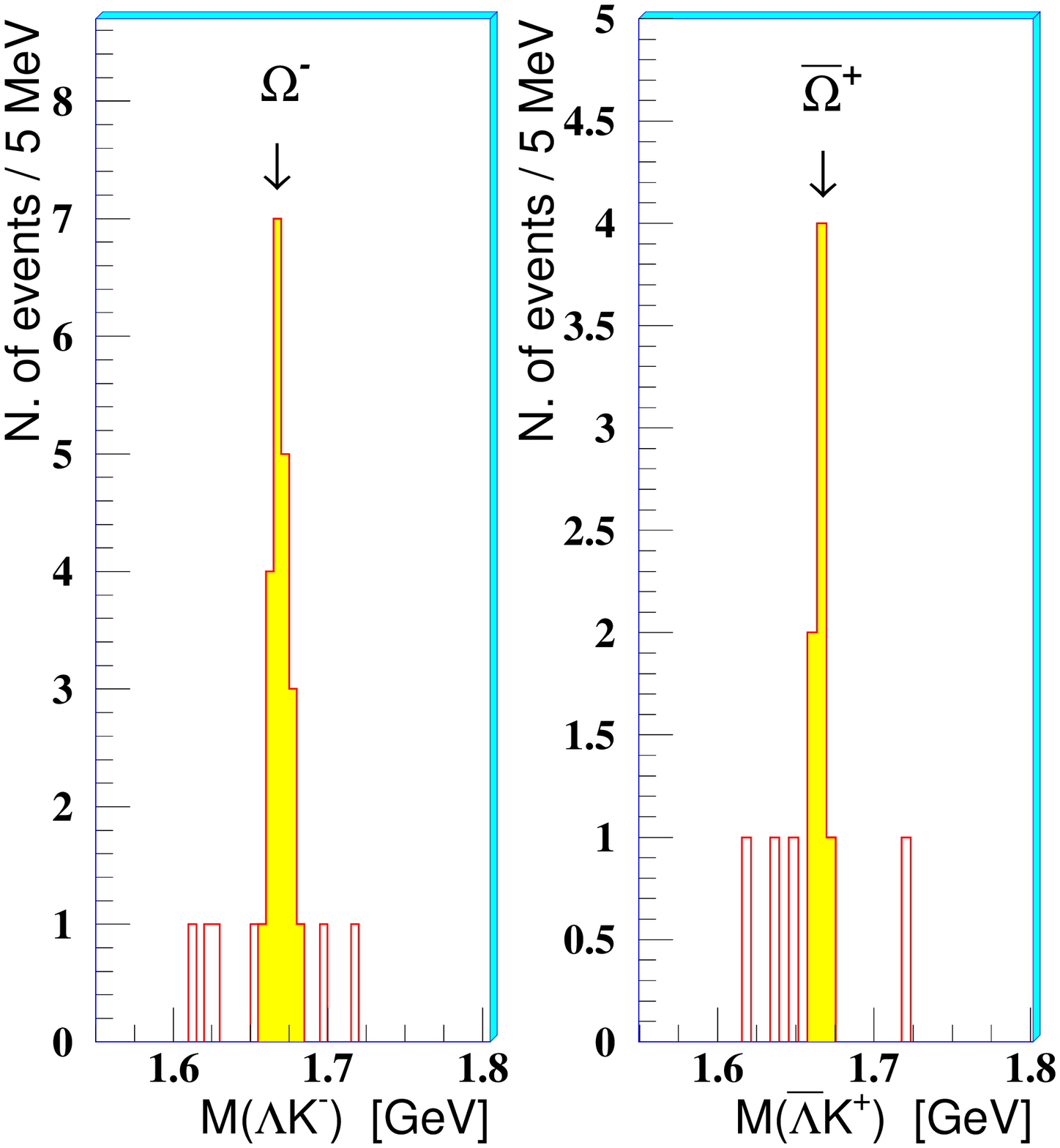}}
\caption{Sample invariant mass spectra for \Pgpp\Pgpm, \Pp\Pgp,
              \PgL\Pgp\ and \PgL\PK. The arrows show the PDG mass 
              values~\cite{PDG}.}
\label{fig:signals}
\end{figure}

The strange particle signals are extracted with the same procedure  
used at 158 $A$\ GeV/$c$~\cite{enh160,BlastPaper}   
by applying similar geometric and kinematic constraints. 
The invariant mass spectra of \Pgpm\Pgpp, \Pp\Pgp, \PgL\Pgp\ and \PgL\PK\  
combinations after all analysis cuts are shown in figure~\ref{fig:signals}.  
The quality of the signals %is quite as good as at higher energy, 
is comparable to that at 158 GeV/$c$~\cite{enh160,BlastPaper}, 
with the hyperon mass peaks at the nominal PDG values~\cite{PDG} and FWHMs of 
about 15 MeV/$c^2$.  
The \PKzS\ invariant mass peak has a FWHM  of 25 MeV/$c^2$\ and its maximum is shifted 
down 
%towards lower values 
by 4 MeV/$c^2$\ with respect to the nominal value. This effect 
%has been understood as due to a slight under-estimation of the total length of 
%pixel telescope (under-steering) and it 
has been accounted for in 
the calculation of the corrections for acceptance and reconstruction inefficiencies.  
The selected particles have been chosen 
%beneath 
%under  
in  
the invariant mass 
%peaks in the 
intervals corresponding to the shaded areas of figure~\ref{fig:signals}.   

\begin{figure}[hbt]
\centering
\resizebox{0.77\textwidth}{!}{%
\includegraphics{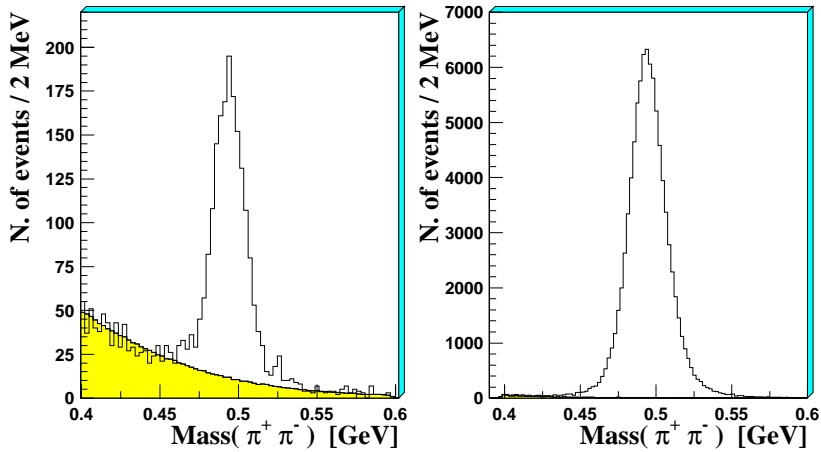}}
\caption{%Comparison between real and mixed events for the
The \Pgpp\Pgpm\ invariant mass distribution
%for real and mixed events
for a small sample of events before analysis cuts (left) and for the total sample 
after analysis cuts (right). The shaded histograms show the combinatorial 
background evaluated by event mixing.}  
\label{fig:Mixing}
\end{figure}

The amount of residual combinatorial background has been evaluated 
%for the high statistics samples (\PKzS, \PgL\ and \PagL) 
using the event-mixing technique~\cite{BlastPaper,BrunoMoriond02}. 
As an example, figure~\ref{fig:Mixing} shows 
the \Pgpp\Pgpm\ invariant mass distribution for real and mixed events before 
(left) and  after (right) the application of the analysis cuts.  
For the cascade hyperons, the residual background has been evaluated to be
about 4\% for \PgXm\  and less than 10\%  for \PagXp\ and \PgOm+\PagOp;  
for the singly-strange particles, it has been 
estimated to be 1\%, 0.8\% and 2\% for \PKzS, \PgL\ and \PagL, respectively.  

%For the cascade hyperons, the residual background has been evaluated to be 
%about 4\% for \PgXm\  and less than 10\%  for \PagXp\ and \PgOm+\PagOp.  

\begin{figure}[hbt]
\centering
\resizebox{0.74\textwidth}{!}{%
\includegraphics{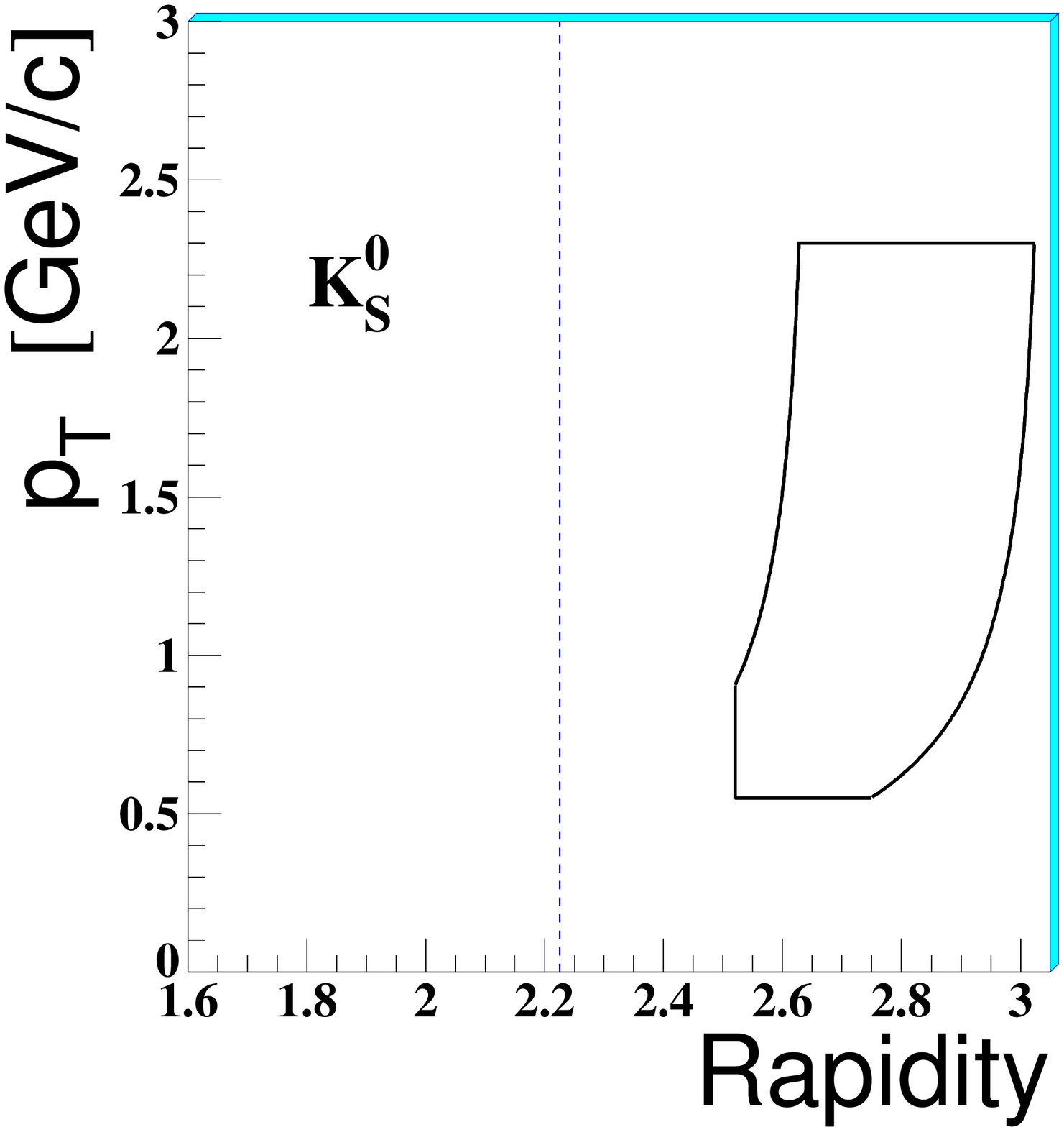}
\includegraphics{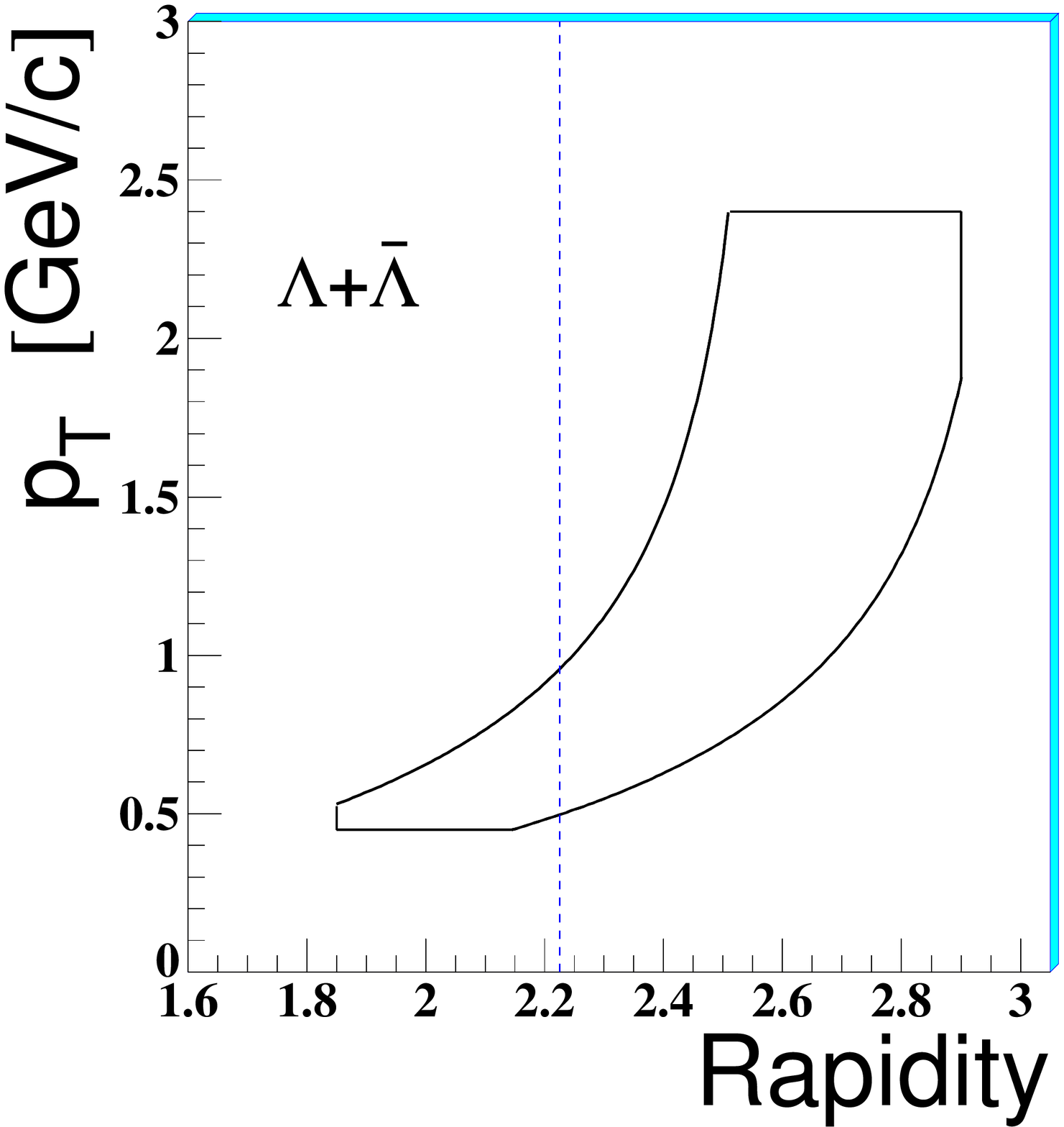}}\\
\resizebox{0.74\textwidth}{!}{%
\includegraphics{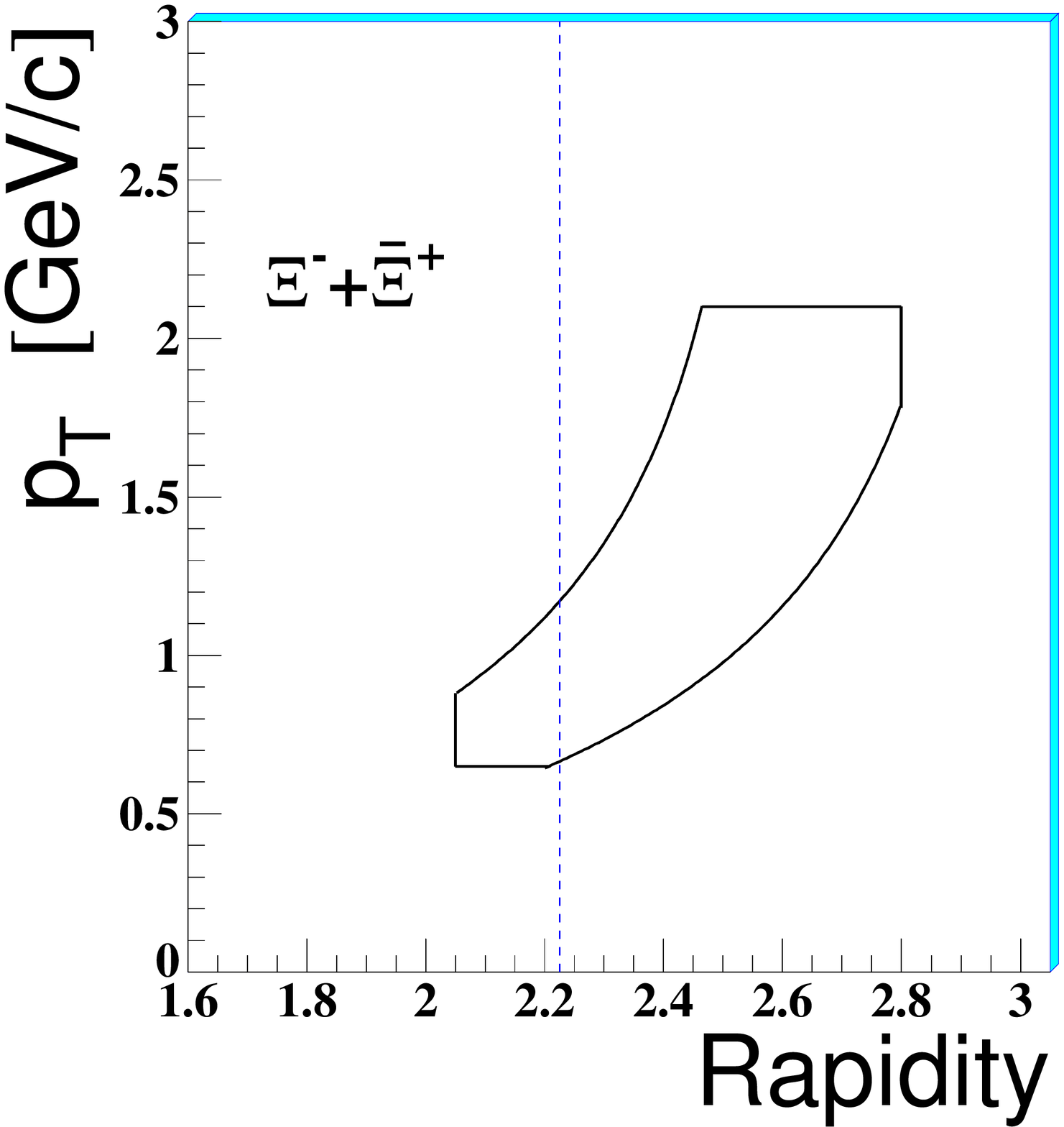}
\includegraphics{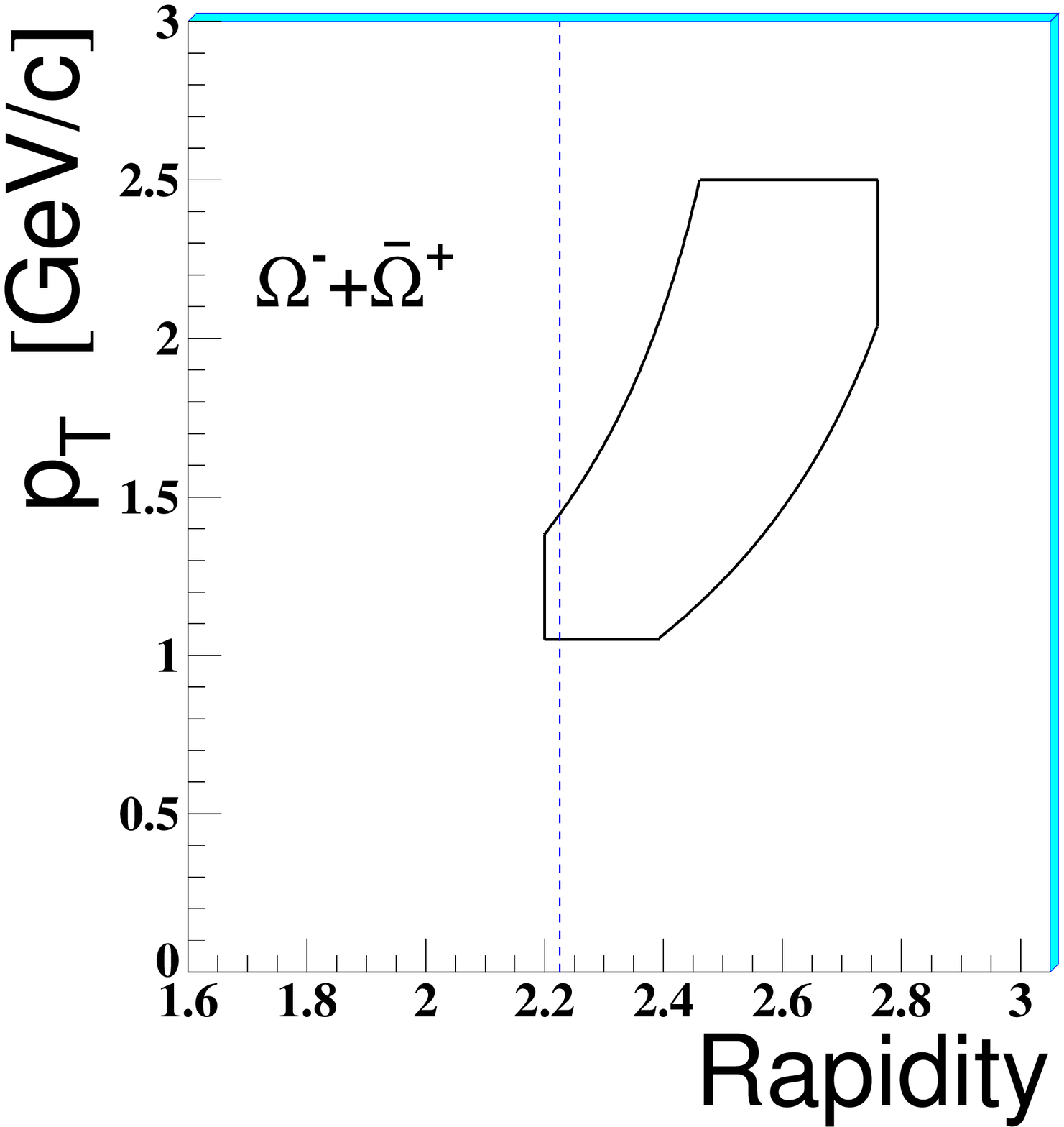}}
\caption{The $\pt$--$y$\ acceptance windows. Dashed lines show the 
         position of mid-rapidity ($y_{\tt cm}=2.225$).}
\label{fig:acceptance} 
\end{figure}

The acceptance regions in the transverse momentum ($\pt$) versus rapidity ($y$) plane 
are shown in figure~\ref{fig:acceptance}. The %exact definitions 
limits  
of these windows have been 
defined in order to exclude from the final sample the strange particles whose 
lines of flight are very close to the 
%border limits 
borders  
of the telescope, where the systematic 
errors are more difficult to %be evaluated.  
evaluate.  

Each reconstructed particle %has 
%needs to 
%must  be 
is  
assigned a weight to correct for  
acceptance and reconstruction inefficiencies. The computational algorithm 
of the event weight is the same used at higher 
energy~\cite{enh160,BlastPaper}: a number of Monte Carlo events are generated --- 
each event consisting of one simulated particle, with the $\pt$\ and $y$\ 
of the real particle and random azimuthal and internal decay angles, merged with 
a real event of similar telescope hit multiplicity as the original event --- and they  
%those events 
are reconstructed with the same analysis tools as for real events;  
the final weight is the ratio of the number of generated particles to 
the number of those reconstructed and retained by the selection criteria.   
This individual correction procedure has been applied to all the reconstructed 
$\Omega$, $\Xi$\ and \PagL\ particles. 
For the much more abundant \PKzS\ and \PgL\ %samples, 
%particles,  
candidates,  
the individual weights 
have been computed for a subsample only.  
These amount to 1/20 and 1/15 of the total sample, respectively, and they    
have been extracted uniformly over the full data taking period.  
The statistics of particles collected and individually corrected is 
given in table~\ref{tab:statistics}. 

\begin{table}[hbt]
\caption{Statistics of the collected and weighted particles.  
\label{tab:statistics}}
\begin{center}
\begin{tabular}{lcccccc}
\hline
       &   \PKzS   &   \PgL   &  \PagL  & \PgXm   &  \PagXp &\PgOm+\PagOp\\ \hline
%total  &   97000  &  82500   &  2100   &  424    &    35   &    28      \\
total  &   97000  &  82500   &  2100   &  439    &    39   &    28      \\
${\rm \frac{weighted}{total}}$  
       &   1/20    &  1/15   &    1    &   1     &     1   &     1       \\\hline
\end{tabular}
\end{center}
\end{table}

The correction procedure has been extensively checked by comparing real and Monte Carlo 
distributions for several parameters --- in 
particular those used for particle selection, such as the distance of closest approach  
in space between the two decay particles, the position of the decay vertices, 
the impact %parameters of the particles at the main vertex position
parameters of the particles\footnote{
%The impact parameter is
%defined as the distance (perpendicular to the beam direction) between the
%extrapolated lines of flight of the particle and the primary vertex position.  
The impact parameter 
is approximated  
%is defined  %, with respect to the primary vertex position, 
%as the intersection of the 
as the distance from the primary vertex of the intersection of the 
measured 
particle trajectory 
with a plane transverse to the beam line passing through the target position.  
}, etc.~--- 
for different data taking periods and magnetic field orientations (up and down). 

In order to further check the stability of the results (i.e. the 
%double-differential  
$\frac{1}{\mt}\frac{\diffD^2N}{\diffD\mt\,\diffD y }$\ %(y,\mt)$\ 
distributions) the selection criteria have also been varied, 
either by changing their limiting values or by excluding them one at a time.  
As a result of these studies we can estimate the contribution of the selection and 
correction procedure to the systematic errors on the slope of the  
$\frac{1}{\mt}\frac{\diffD N}{\diffD\mt}$\  
distributions to be about 12\% for $\Omega$\ and 8\% for all other particles.  

%%
%% Other sources to be mentioned: residual background, shape of the $y$ 
%% spectra

The experimental procedure for the determination of the $m_{\tt T}$\ 
distribution is described in detail in reference~\cite{BlastPaper}, 
where the results
%for Pb--Pb collisions
at 158 $A$\ GeV/$c$\ are discussed.   
The measured distribution of the double differential invariant cross-section
$\frac{1}{\mt}\frac{d^2N}{\diffD\mt \, \diffD y} $\ has been assumed to factorize into  
a $y$\ and an $\mt$\ %(or $\pt$) 
dependent part:    
\begin{equation}
\label{eq:factorize}
\frac{1}{\mt}\frac{\diffD^2N}{\diffD\mt\,\diffD y }(y,\mt) = f(y) \cdot 
    \frac{1}{\mt}\frac{ \diffD N }{\diffD \mt} (\mt)   .
\end{equation}
This assumption has been verified by considering 
the  $\mt$\ ($y$) distributions for different slices in rapidity 
(transverse mass). %: the shapes of the distributions stay constant 
%within less than 10\% for all the particles. 

%As at 158 $A$\ GeV/$c$\ beam momentum~\cite{RapPaper},  
The shape of the rapidity distribution, i.e. $f(y)$\ in equation~\ref{eq:factorize},   
has been found to be well described %within our limited range 
within our limited acceptance  
by a constant for all particles except for \PKzS\ and \PagL, where a Gaussian 
provides a better description.  
We have reported 
a similar finding %has been reported 
in Pb--Pb collisions at 158 $A$\ GeV/$c$\ beam momentum~\cite{RapPaper}. 

The hypotheses on the factorization of the double differential invariant 
cross-section  (equation~\ref{eq:factorize}) and on the shape of the 
rapidity distributions ($f(y)$) can introduce a systematic bias on   
the slope of the $\frac{1}{\mt}\frac{ \diffD N }{\diffD \mt} (\mt) $\  
distributions which has been estimated not to exceed 5\%.  

%contribution of systematic errors on the slope of the 
%$\frac{ \diffD N }{\mt \diffD \mt} (\mt) $\ 
%distributions which has been estimated %to be about 5\%.  
%not exceeding 5\%.  

An additional source of systematic errors comes from the residual 
combinatorial background, which can have a different $\mt$\ distribution than the signal.    
This error is found to be negligible for \PgL\ and \PKzS, and about 1.5\%, 2\%, 
4\% and 7\%  for $\PagL$, $\PgXm$, $\PagXp$ and $\Omega$, respectively.

\section{Exponential fits of the transverse mass spectra}  
\begin{figure}[hbt]
\centering
\resizebox{0.74\textwidth}{!}{%
\includegraphics{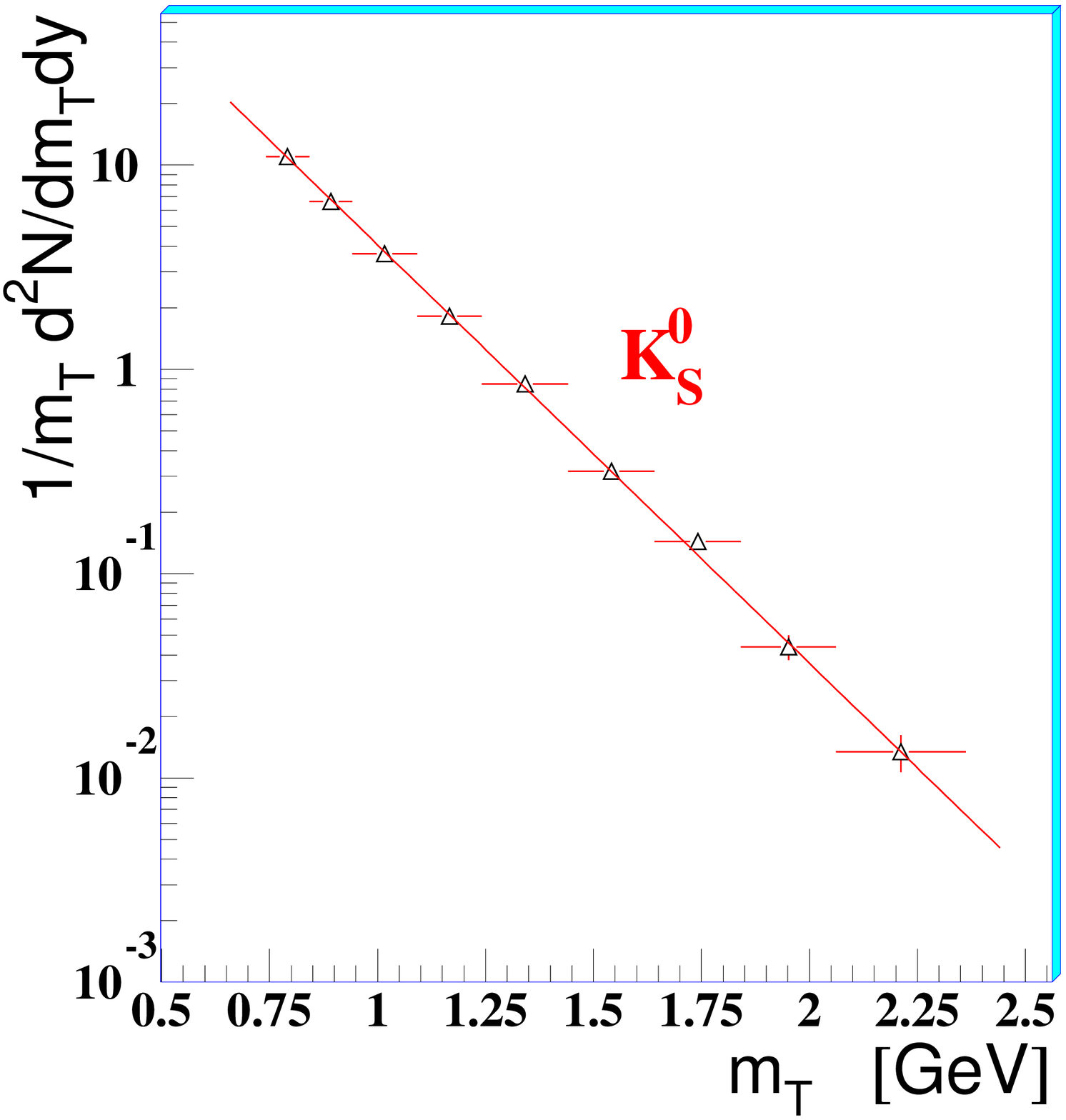}
\includegraphics{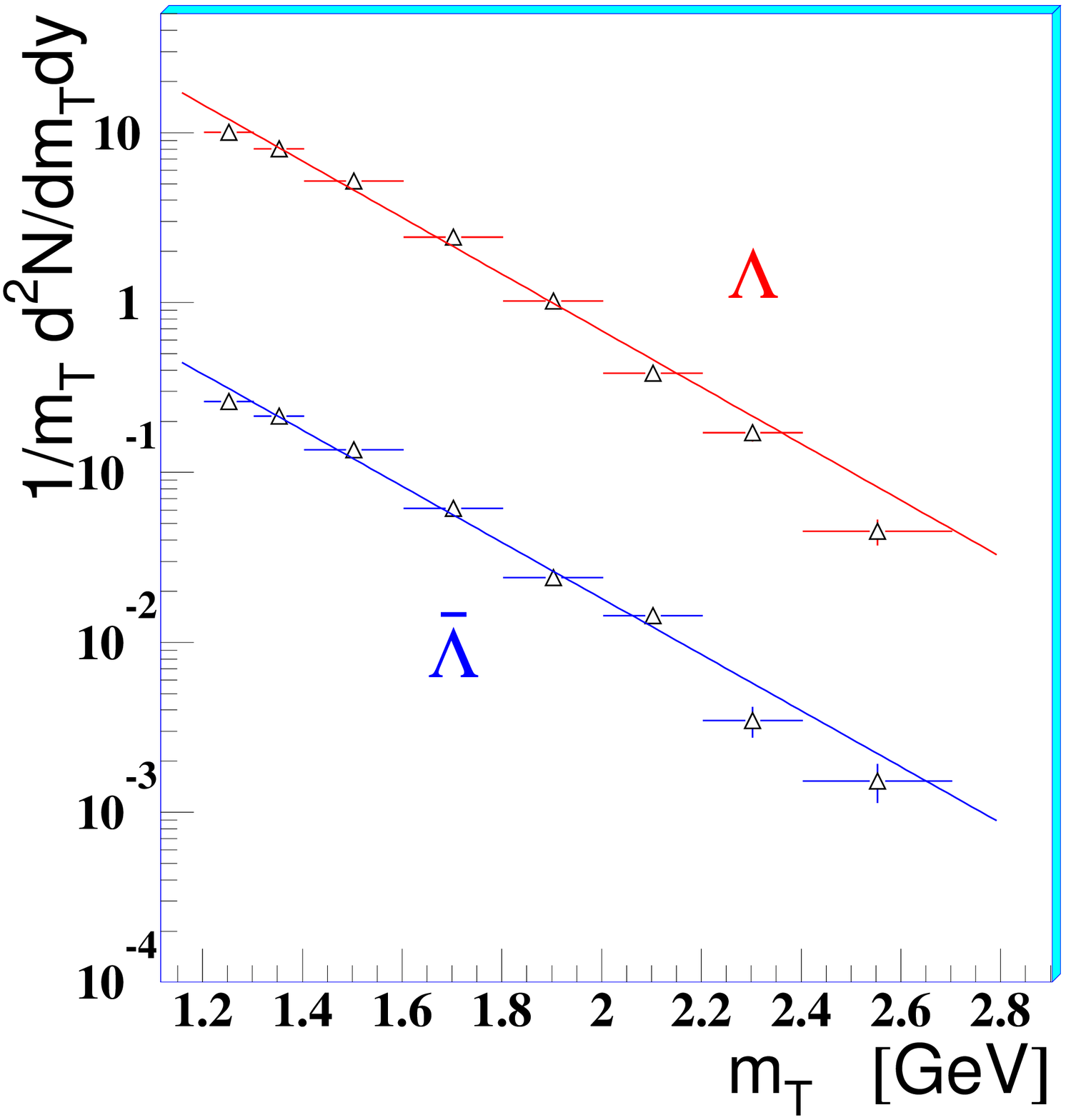}}\\
\resizebox{0.74\textwidth}{!}{%
\includegraphics{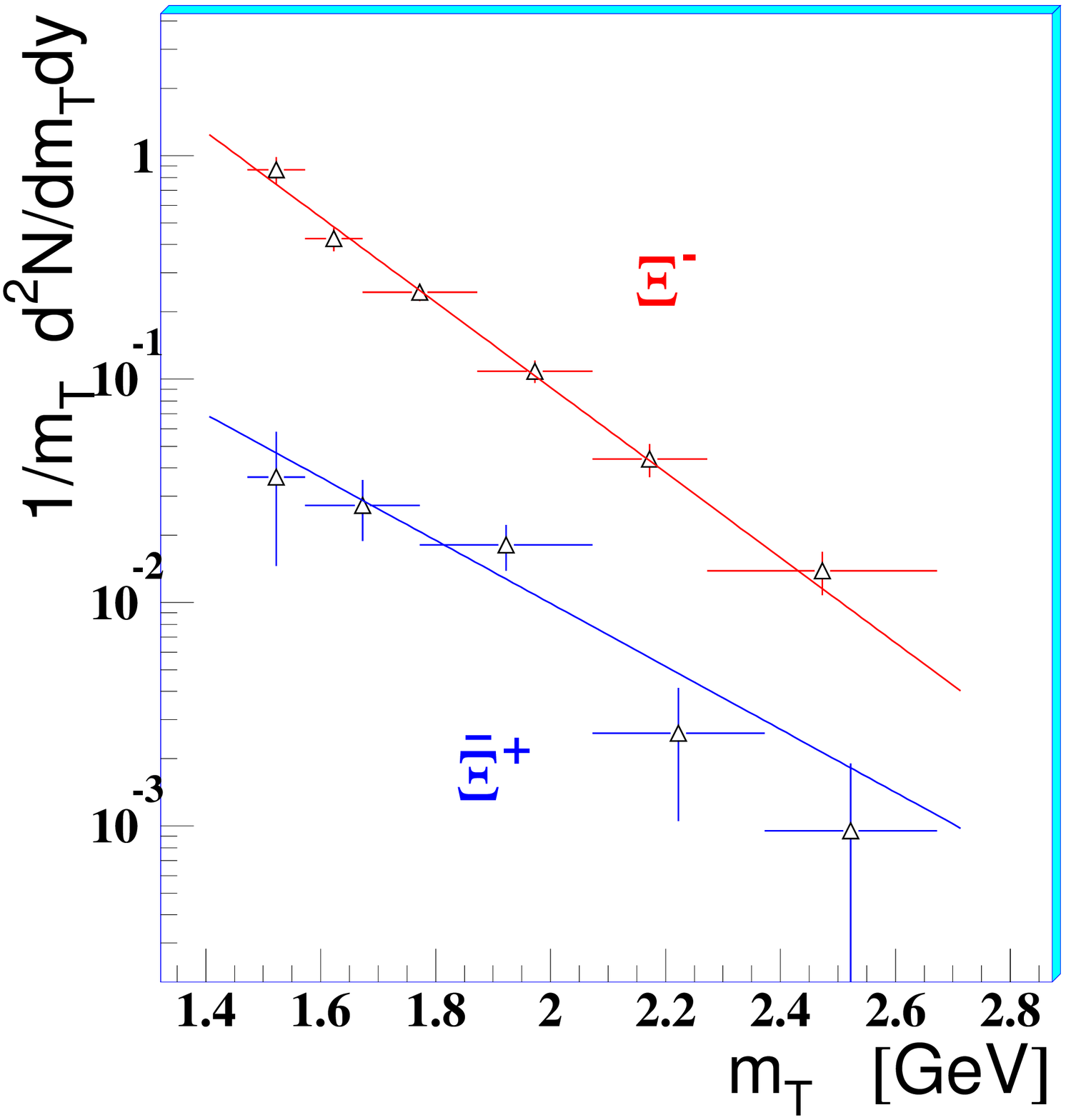}
\includegraphics{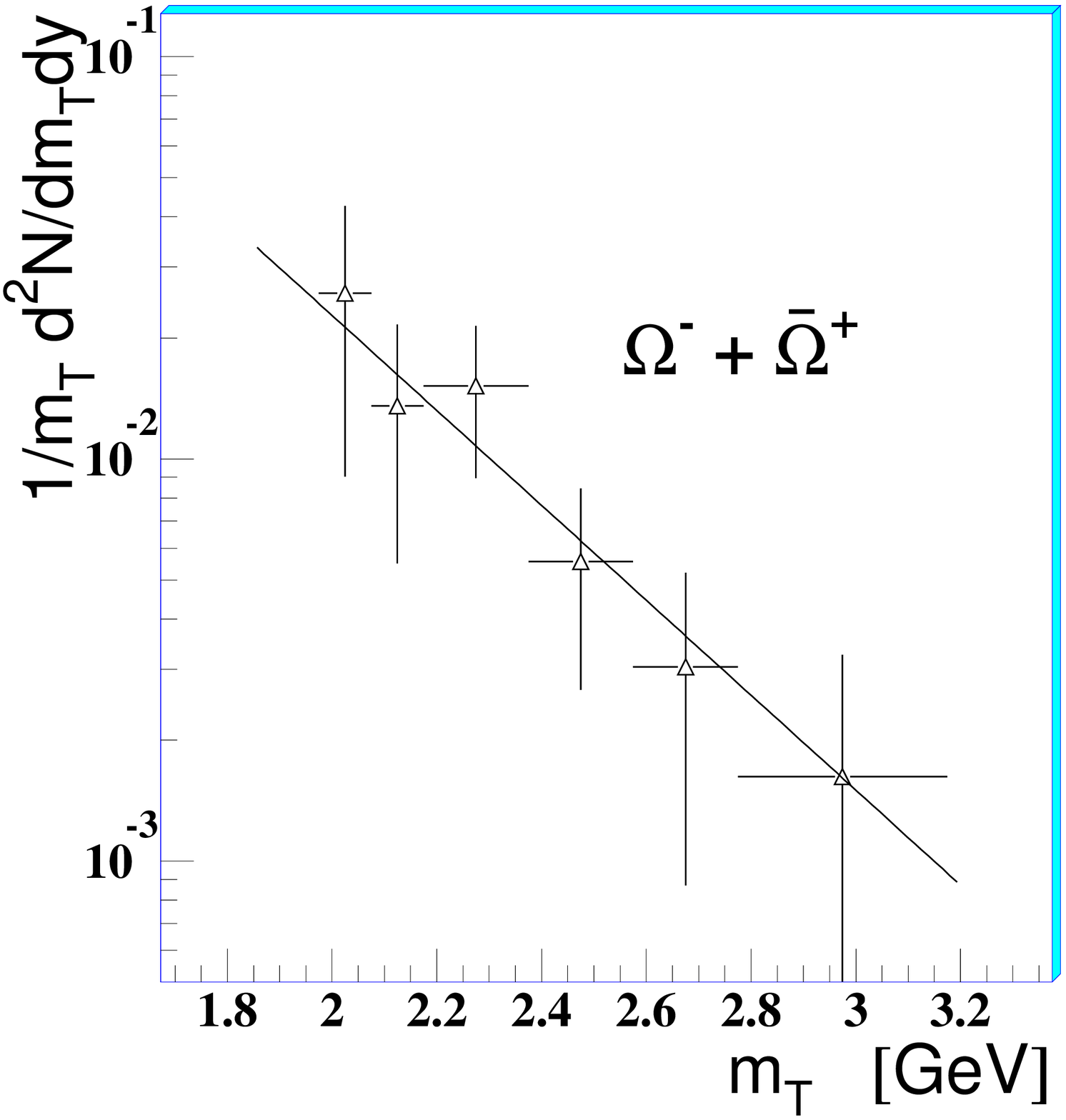}}
\caption{ Transverse mass spectra of strange particles
 for the most central 53\% of the Pb--Pb inelastic cross-section.
 The superimposed exponential functions have inverse slopes equal to the
 $T_{\tt app}$\ values obtained from the maximum likelihood fits.}
\label{fig:all_spectra}
\end{figure}

The inverse slope parameter $T_{\tt app}$\ (``apparent temperature'')  
has been extracted by means of a maximum likelihood fit 
%of equation~\ref{eq:expo} 
of the measured double differential invariant cross-section 
$\frac{1}{\mt}\frac{d^2N}{\diffD \mt \, \diffD y} $ 
to the formula  
\begin{equation}
\frac{1}{\mt}\frac{d^2N}{\diffD \mt \, \diffD y} =
   f(y) \hspace{1mm} \exp\left(-\frac{m_{\tt T}}{T_{\tt app}}\right) 
\label{eq:InvSlope}
\end{equation}
%\footnote{The rapidity distribution is assumed to be flat within our 
%acceptance region corresponding to about half a unit of rapidity around mid-rapidity.}.  
The apparent temperature is interpreted as due to the  
thermal motion coupled with a collective transverse flow  
of the fireball components~\cite{BlastRef}.    
The differential invariant cross-section distributions are shown in 
figure~\ref{fig:all_spectra} as a function of $\mt$\ 
for the most central 53\% of the inelastic Pb--Pb cross-section 
with the likelihood fit results superimposed.   
The values of the inverse slope parameters $T_{\tt app}$\ are 
given in table~\ref{tab:InvSlopes}. 
The quoted systematic errors are obtained by propagating the %aforementioned 
partial contributions due to i)~the selection and correction 
procedure, ii)~the hypothesis on the parameterization of the 
double differential cross-section and iii)~the residual combinatorial background.  
\begin{table}[htb]
\caption{Inverse slope parameter $T_{\tt app}$\ (MeV) of the strange particles
in the full centrality range ({\bf 0--53\%}). The first error is statistical, the
second one systematic.  
\label{tab:InvSlopes}}
\begin{center}
\begin{tabular}{|c|c|c|}
\hline
           {\bf $K_S^0$}  &   {\bf $\La$}  &  {\bf $\Al$}   \\ \hline
           $212\pm3\pm21$ & $261\pm4\pm26$ & $263\pm6\pm26$ \\ \hline
\end{tabular}
                                                                                                                             
\begin{tabular}{|c|c|c|} \hline
     {\bf $\Xi^-$}  & {\bf $\overline\Xi^+$} & {\bf $\Omega^- + \overline\Omega^+$} \\ \hline
%     $228\pm12\pm23$ & $308\pm63\pm23$ & $368\pm120\pm40$  \\ \hline
    $228\pm12\pm23$ & $308\pm63\pm31$ & $368\pm120\pm40$  \\ \hline
\end{tabular}
\end{center}
\end{table}
%

%The similarity of the shapes of the \PgL\ and \PagL\ $\mt$\ distributions  
%is %impressive; 
%remarkable;  
%this is reflected in their apparent temperature values 
%which agree within 2\%
The \PgL\ and \PagL\ distributions have very similar inverse slopes, agreeing 
within 2\%\footnote{The systematic  
errors do not play a role in such a comparison since they essentially cancel out 
in the ratio of the spectra. The only different contribution 
which can be envisaged arises from the different %amount of the 
residual background. This has been estimated to affect by less than 1.5\% 
the inverse slope of the \PagL, as mentioned before.}. A similar 
baryon--anti-baryon symmetry was reported at top SPS  
energy for central and semi-central Pb--Pb collisions 
%(over the centrality range 0--40\%, corresponding to the classes from 1 to 4 of table~\ref{tab:centrality}) 
for \PgL\ and  $\Xi$~\cite{BlastPaper}. 
%It was clearly not present in pBe and hardly seen in pPb and peripheral Pb--Pb collisions.  
%The similarity of baryon and antibaryon $m_{\tt T}$\ slopes  
%suggests that strange baryons and antibaryons may be produced and 
%evolve in the collision dynamics by similar mechanisms. 
The similarity of baryon and anti-baryon $m_{\tt T}$\ slopes is 
interpreted as %a suggestion 
suggestive  
that strange baryons and anti-baryons would be 
produced and evolve in the collision dynamics by similar mechanisms. 
Based on the \PgL\ and  \PagL\ results presented above, a similar conclusion can be 
extended to the 40 $A$\ GeV/$c$\ collisions.  
The large statistical errors associated 
to the \PagXp\ $\mt$\ distribution prevent any such  
conclusion for this multi-strange baryon.   

In the hydro-dynamical view, the apparent temperature 
%is a local function of the 
%% $\mt$\ value of the 
%invariant $\mt$\ distribution
depends on $\mt$\footnote{In this view, 
the graphical interpretation of $T_{\tt app}$\ would be the inverse of the 
%local 
tangent to the invariant $1/\mt \, \diffD N / \diffD \mt$\ distribution.  
See reference~\cite{BlastPaper} for a detailed discussion.}.  
At a given %$\overline{\mt}$\ 
${\mt}_0$\  
value, it can be calculated according to the  
formula~\cite{BlastRef}:  
\begin{equation}
T_{\tt app}({\mt}_0)=\left[\lim_{\mt \rightarrow {\mt}_0}
\frac{\diffD}{\diffD\mt}(\ln \frac{\diffD N}{\diffD \mt^2})  \right]^{-1}
\label{eg:Tapp}
\end{equation} 
With some approximations, this expression simplifies for two asymptotic cases: 
at low $\pt$, i.e. $\mt \rightarrow m_0$, it provides  
$T_{\tt app}=T+\frac{1}{2}m_0 \Bt^2 $~\cite{BlastRef,SecondRef}, 
where $T$\ is the freeze-out temperature and $\Bt$\ is the average 
transverse flow velocity;   
at high  $p_t$, i.e. $\mt \rightarrow \infty$, the apparent temperature 
is simply blue-shifted by the collective dynamics, independently of 
the particle mass~\cite{BlastRef}:  
\begin{equation}
T_{\tt app}= T \sqrt{\frac{1+\Bt}{1-\Bt}}.  
\label{eq:BlueShift}
\end{equation}  
Therefore, the apparent temperature actually depends on the 
$\mt$\ range where the $1/\mt \, \diffD N/ \diffD\mt$\ distribution 
has been fitted to an exponential function.  
Nevertheless, provided that the spectra are measured at intermediate $\pt$\ over 
similar $\mt-m_0$\ ranges, 
the inverse slopes should follow a hierarchy with the rest mass of the particles.  
The apparent temperature of the \PgXm\ hyperon is found to be smaller than those of 
the \PgL\ and \PagL\ particles, and of the same order as that of the \PKzS\ meson. 
A similar ``violation'' of the mass hierarchy was reported for the 
triply-strange $\Omega$\ particle by the WA97~\cite{MtWA97}, NA57~\cite{BlastPaper} 
and NA49~\cite{NA49Omega} experiments at 158 $A$\ GeV/$c$, and  was 
interpreted~\cite{Hecke} as an indication of an earlier decoupling of that particle 
from the expanding fireball. In the next section we shall discuss in more details  
the possibility of a similar effect for the \PgXm\  at 40 $A$\ GeV/$c$.  

\begin{figure}[t]
\centering
\resizebox{0.54\textwidth}{!}{%
\includegraphics{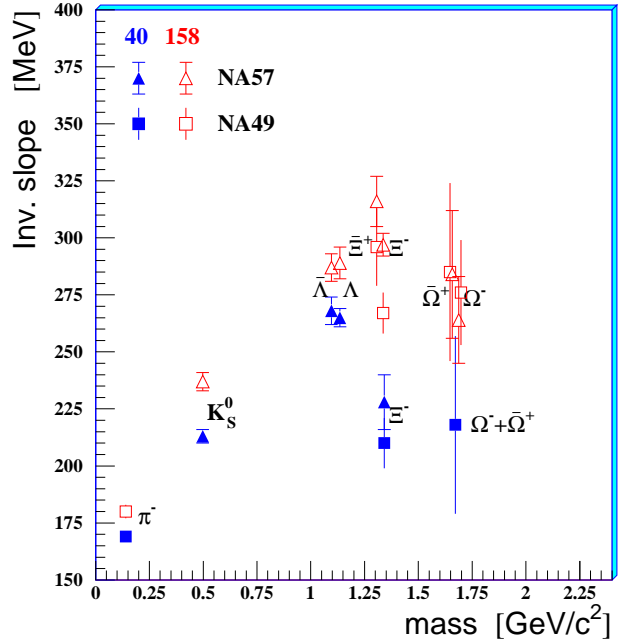}}
\caption{Apparent temperature as a function of the particle rest mass in Pb--Pb 
         at 40 $A$\ GeV/$c$\  (closed symbols, in blue on-line) as  
         compared to 158 $A$\ GeV/$c$ (open symbols, in red).  
         NA49 results for negative pions~\cite{NA49pions} and multi-strange 
         hyperons~\cite{NA49Xi,NA49Omega,NA49Xi40} are also shown (squares).   
         For display purpose the ordinate scale has been zero-suppressed. 
         %\newline
         %{\em Note: NA49 \PgXm\ at 40 GeV are preliminary.} 
           }
\label{fig:InvSlopVsEnergy}
\end{figure}

Figure~\ref{fig:InvSlopVsEnergy} shows a comparison of the inverse slopes measured 
at the two energies   
plotted as a function of the particle rest mass.  
The inverse slopes are lower at lower energy by about 7\% for non-strange and 
singly-strange particles and about 20\% for the multi-strange ones.

\subsection{Centrality dependence}
\begin{figure}[t]
\centering
\resizebox{0.84\textwidth}{!}{%
\includegraphics{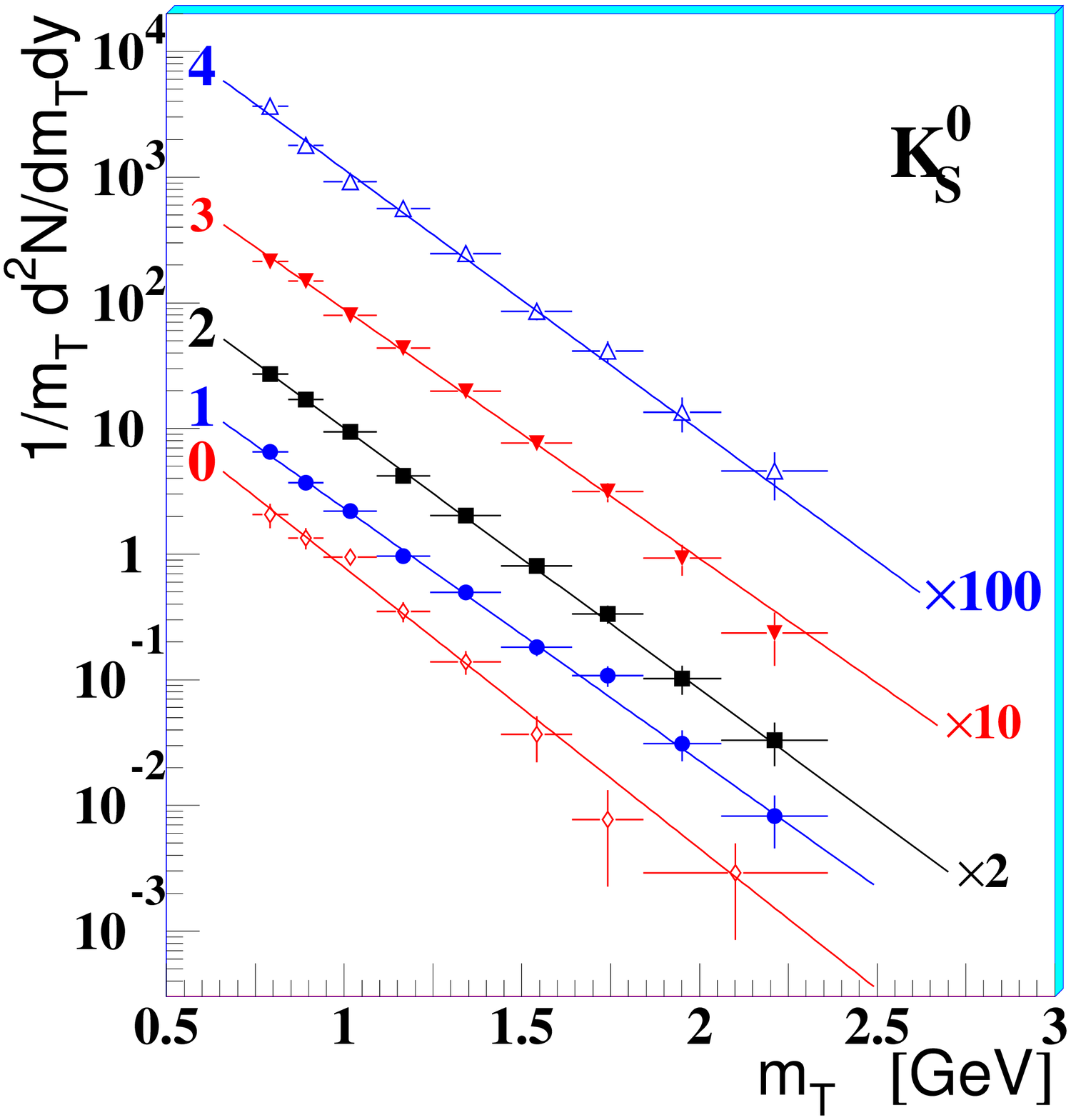}
\includegraphics{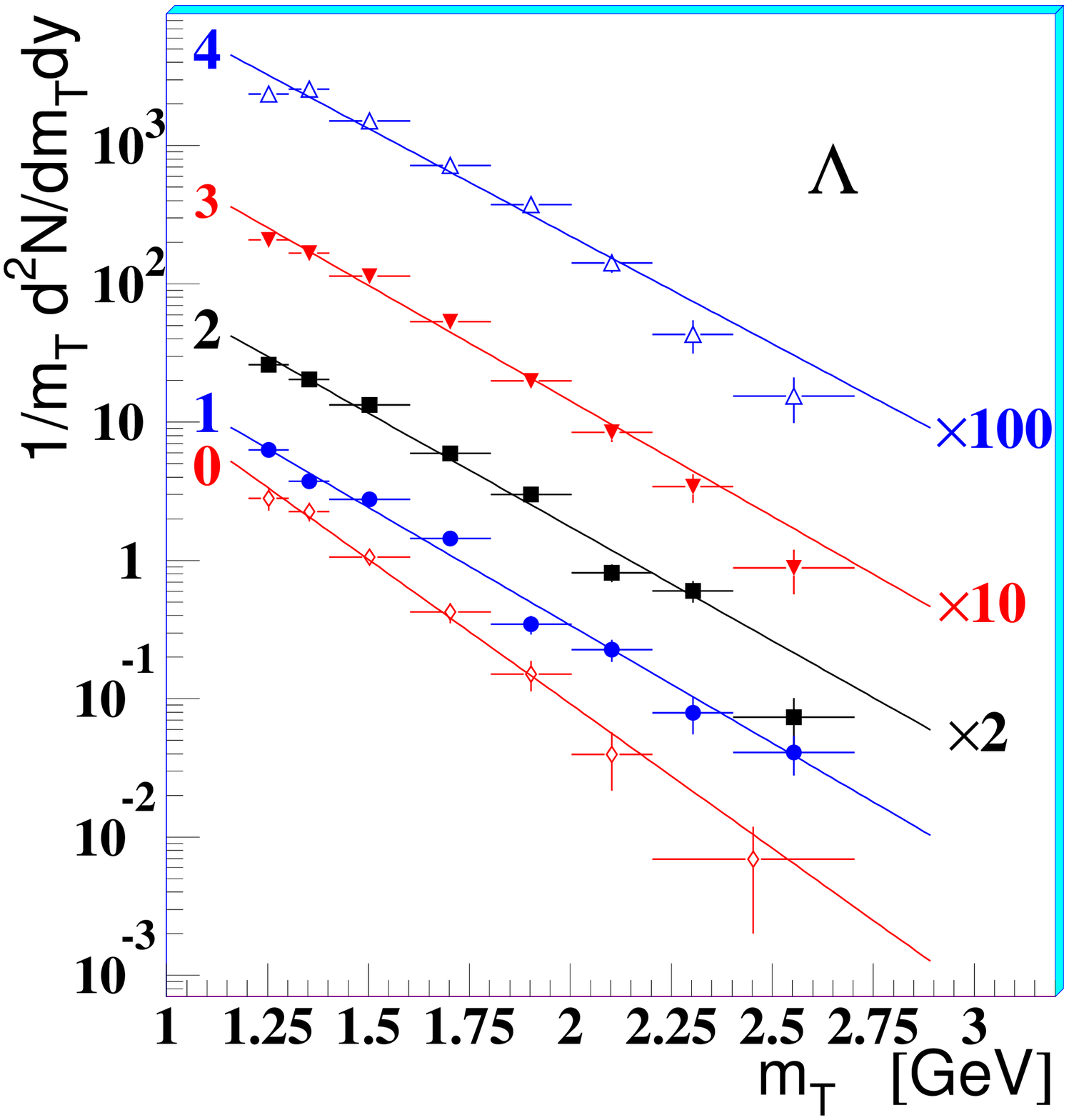}}\\
\resizebox{0.84\textwidth}{!}{%
\includegraphics{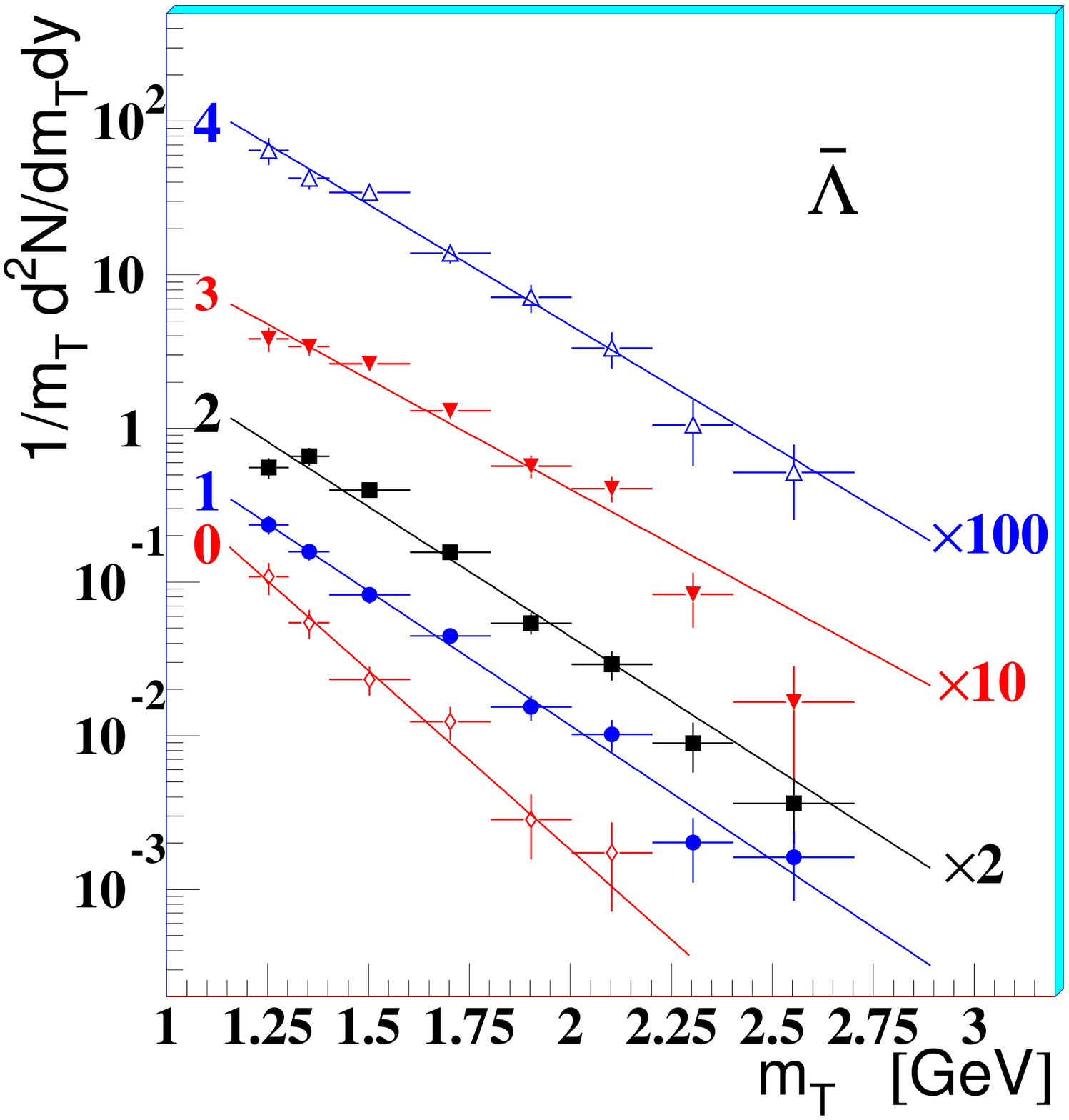}
\includegraphics{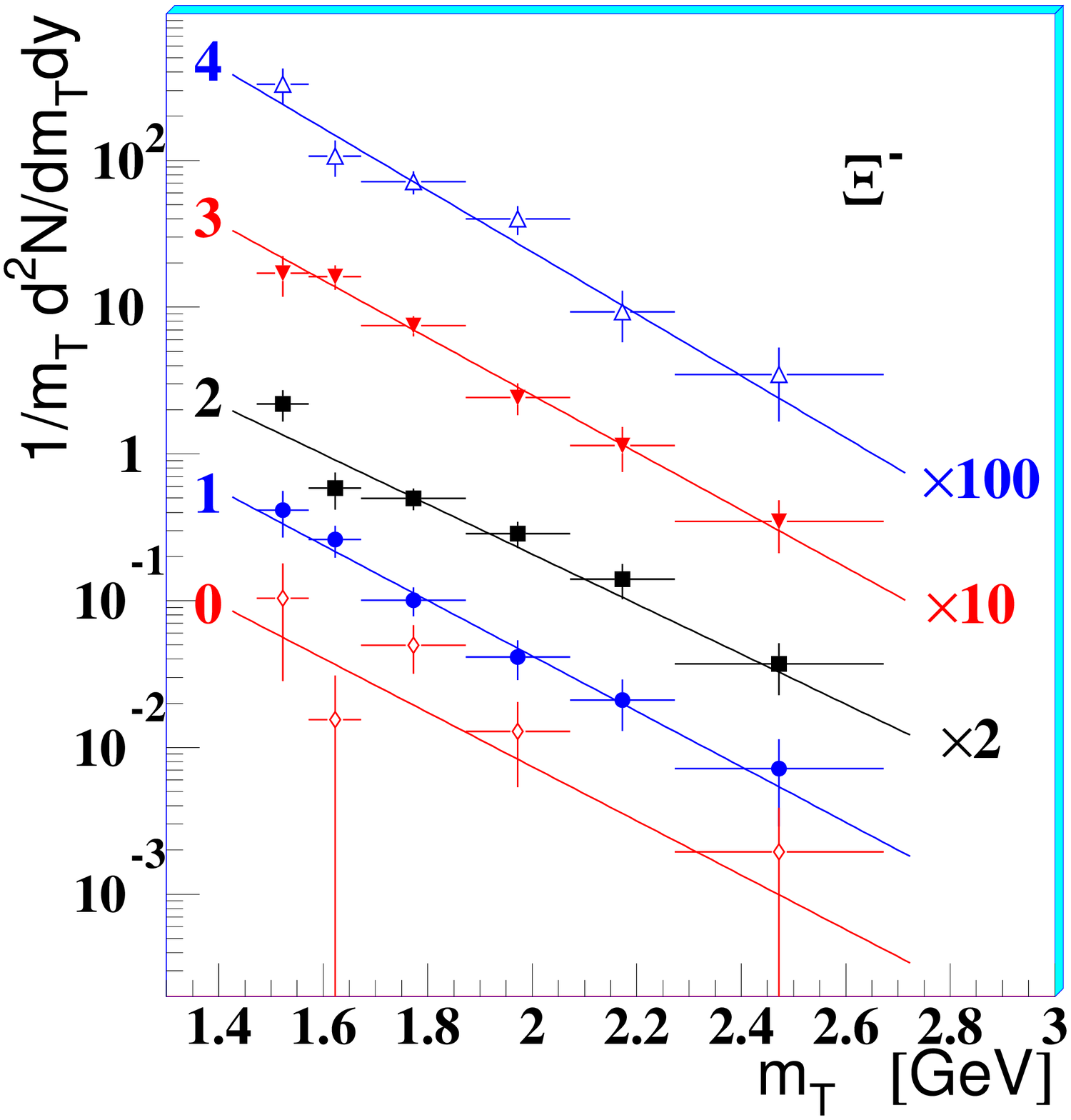}}
\caption{  Transverse mass spectra of \PKzS, \PgL, \PagL\ and \PgXm\  
           in Pb--Pb collisions at 40 $A$\ GeV/$c$\ for
           the five centrality classes of table~\ref{tab:centrality}.
           For each species, class $4$\ is displayed uppermost and
           class $0$\ lowermost. The spectra 
           of class $2$, $3$\ and $4$ have been scaled by factor $2$, $10$\ 
           and $100$, respectively, for display purposes. 
	   }
\label{fig:msd_spectra}
\end{figure}

The transverse mass spectra measured in the individual centrality classes of 
table~\ref{tab:centrality} are shown in figure~\ref{fig:msd_spectra} 
for \PKzS, \PgL, \PagL\ and \PgXm.  For the rarer \PagXp\ and $\Omega$\ hyperons, 
the collected statistics do not allow to study the centrality dependence.
Maximum likelihood exponential fits are 
superimposed to the spectra, as in figure~\ref{fig:all_spectra}.  
The inverse slope parameters $T_{\tt app}$\ %of the transverse mass spectra 
are given  in table~\ref{tab:InvMSD} as a function of centrality.  
\begin{table}[h]
\caption{
         Inverse slopes (MeV) of the  
	 %strange particle  
         $m_{\tt T}$\ distributions of \PKzS, \PgL, \PagL\ and \PgXm\ 
         as a function of centrality at 40 $A$\ GeV/$c$.  
	 Only statistical errors are shown. Systematic errors are 
	 estimated to be 10\% for all centralities.   
\label{tab:InvMSD}}
\begin{center}
%%%%%%%%%%%%%%%%% NEW with pBe %%%%%%%%%%%%%%%%%%%%%%%
\footnotesize{
\begin{tabular}{|c|c|c|c|c|c|} 
\hline
      &     0      &    1      &    2      &    3    &    4   \\ \hline
\PKzS & $194\pm12$ & $216\pm7$ & $209\pm5$ & $219\pm6$ & $209\pm7$ \\
\PgL  & $208\pm12$ & $255\pm9$ & $264\pm7$ & $260\pm7$ & $279\pm9$ \\
\PagL & $186\pm17$ & $248\pm12$& $257\pm11$& $303\pm15$& $276\pm16$ \\
\PgXm & $235\pm67$ & $230\pm30$ &$255\pm26$& $222\pm20$& $206\pm22$ \\
\hline
%%%%%%%%%%%%%%%%%%%%%%%%%%%%%%%%%%%%%%%%%%%%%%%%%%%%%%%
\end{tabular}
}
\end{center}
\end{table}

%A significant 
An  
increase of the apparent temperature when going from 
class 0 to class 1 is observed 
for the singly-strange particles. Then, from class 1 to 4, the inverse slopes of 
these particles remain approximately constant. 
%with  a possible weak increase with centrality for the \PgL. %which cannot be excluded from data. 
The baryon--anti-baryon symmetry in the shape of the spectra discussed 
above for the \PgL\ hyperon is preserved also as a function of centrality.  

No centrality dependence is observed within the errors 
for the \PgXm\ hyperon.  
Remarkably, the apparent temperature of the \PgXm\ hyperon is below those 
of \PgL\ and \PagL\ and compatible with that of the light \PKzS\ meson  
for all centrality classes apart for class 0 where the mass hierarchy 
appears to be reestablished.  
\section{Blast-wave description of the spectra}
In this section we employ the statistical hadronization model of 
reference~\cite{BlastRef}, which has provided a good description of  
the 158 $A$\ GeV/$c$\ results~\cite{BlastPaper,RapPaper},  
to study the strange particle $\mt$\ spectra discussed above.   
%as we did in the  
%previous analysis of the 158 $A$\ GeV/$c$\ data~\cite{BlastPaper}.  
%This model provided a nice description of the 158 $A$\ GeV/$c$\ data~\cite{BlastPaper}
The model assumes that particles decouple from a system in local thermal  
equilibrium with a temperature $T$, and which expands both 
longitudinally   
and in the transverse direction;    
the longitudinal expansion is taken to be boost-invariant  
and the transverse expansion is defined in terms of a transverse velocity 
profile. 
%Finally, the statistical distributions are   
%approximated by the Boltzmann distribution.  
%\newline
In this model  
%The blast-wave model~\cite{BlastRef} predicts 
the differential cross-section    
for particle $j$\  has  
%($j=\PKzS,\PgL, \PagL, \PgXm, \PagXp, \PgOm, \PagOp$) 
the form:  
\begin{equation}
\frac{1}{\mt}\frac{\diffD^2N_j}{\diffD \mt \diffD y} %\propto 
    = \mathcal{A}_j  \int_0^{R_G}{ 
     m_{\tt T} K_1\left( \frac{m_{\tt T} \cosh \rho}{T} \right)
         I_0\left( \frac{p_{\tt T} \sinh \rho}{T} \right) r \, {\rm d}r}
\label{eq:Blast}
\end{equation}
where $\rho(r)=\tanh^{-1} \beta_{\perp}(r)$\ is a transverse boost,   
$K_1$\ and $I_0$\ are  modified Bessel functions, $R_G$\ is the 
transverse geometric radius of the source at freeze-out 
and $\mathcal{A}_j$\ is a normalization constant.  
%With respect to a cylindrical reference system ($r$,$\phi$,$z$,$t$), 
%the freeze-out hypersurface is constrained by $0 \le r \le R_G$,   
%$ 0 \le \phi \le 2\pi$ and $\partial t_f / \partial r = 0$;   
%the last condition assumes 
%It is assumed 
%that the particles decouple suddenly   
%from  the whole transverse profile of the 
%fireball at time $t_f$.   
%\newline
%In case of a peripheral collision, the azimuthal symmetry is evidently broken; 
%however it is recovered when considering the $m_{\tt T}$\ spectrum of particles 
%accumulated over many events with random impact parameters, which is 
%our approach.   
% 
The transverse velocity profile $\beta_{\perp}(r)$\ has been parameterized as  
\begin{equation}
\beta_{\perp}(r) = \beta_S \left[ \frac{r}{R_G} \right]^{n}  
  \quad \quad \quad r \le R_G
\label{eq:profile}
\end{equation}  
%With this type of profile 
The numerical value of $R_G$\ does not 
influence the shape of the spectra but just the absolute  normalization 
(i.e. the constant $\mathcal{A}_j$).  
%
%Once the transverse flow profile (i.e. equation~\ref{eq:profile}) has been fixed,  
%the shape of each spectrum is determined by the temperature,
%the velocity of the transverse expansion and the mass of the particle.   
The parameters which can be extracted from the fit of equation~\ref{eq:Blast} to 
the experimental spectra are thus the thermal freeze-out 
temperature $T$\ and the surface
transverse flow velocity $\beta_S$. 
%In order to compare results from different profile hypotheses, 
%corresponding to different values of the exponent $n$\ in equation~\ref{eq:profile}, 
%the average transverse flow velocity can be used instead. 
Assuming a uniform particle density, the 
latter can be %connected 
related  
to the {\em average} transverse flow 
%velocity~\cite{Nonso},   
velocity using the expression
%the average\footnote[2]{A more sophisticated averaging can be 
%achieved by incorporating not only the transverse geometry of the model 
%but also the phase space density of particles~\cite{Nonso}.  
%According to this definition, $<\beta_{\perp}>$\ is also  
%a function of $T$\ and it differs from the  
%values calculated according to equation~\ref{eq:averageB} 
%by 2\% if $n=1/2$, by 5\% if $n=1$\ and by 10\% if $n=2$; obviously 
%for $n=0$, $<\beta_{\perp}>=\beta_S$\ independently of the average definition.}  
%\begin{equation}
$
\Bt = \frac{2}{2+n}  \beta_S
$.   
%
%%% Part on limited rapidity. 
%

%The parameterization of equation~\ref{eq:Blast}, 
Equation~\ref{eq:Blast},  
which is obtained by integrating the Cooper--Frye invariant distribution function~\cite{35}  
over the rapidity of fluid elements ($\eta$) up to a maximum longitudinal flow  
$\eta_{\tt max}$~\cite{BlastRef}, is a good approximation of a full hydro-dynamical calculation  
%untill the experimental rapidity coverage about central rapidity is small.  
for small rapidity windows about mid-rapidity.  
When such a hypothesis is not fulfilled, 
a %tedious 
numerical integration should be performed which, however, requires {\em a priori} knowledge of the maximum longitudinal 
flow\footnote{The $\eta_{\tt max}$\ can be derived from a fit to the ${\rm d}N/{\rm d}y$\ distributions 
as shown, e.g., in reference~\cite{RapPaper}.}.  
The smaller the values of the longitudinal flow, the larger the deviations from equation~\ref{eq:Blast}.
%The deviations from equation~\ref{eq:Blast} increase for small values of the longitudinal flow.  
%should be performed when such a hypothesis is not fulfilled. 
In order to estimate the maximum %systematic errors 
bias introduced by this approximation, 
we have performed the integration by assuming that the 
longitudinal flow has the same strength as the transverse one, whereas the collective 
expansion is expected to be stronger in the longitudinal direction due to the 
incomplete stopping of the incoming nucleons.   
%and the larger the longitudinal flow the better the approximation of equation~\ref{eq:Blast}.  
Both freeze-out parameters $T$\ 
and $\Bt$\ are found to be 
smaller by 2\% when the full integration is performed.    
%reduced by 2\% with the use of 

\subsection{Global fit with different profiles}
\begin{figure}[tb]
\centering
\resizebox{0.80\textwidth}{!}{%
\includegraphics{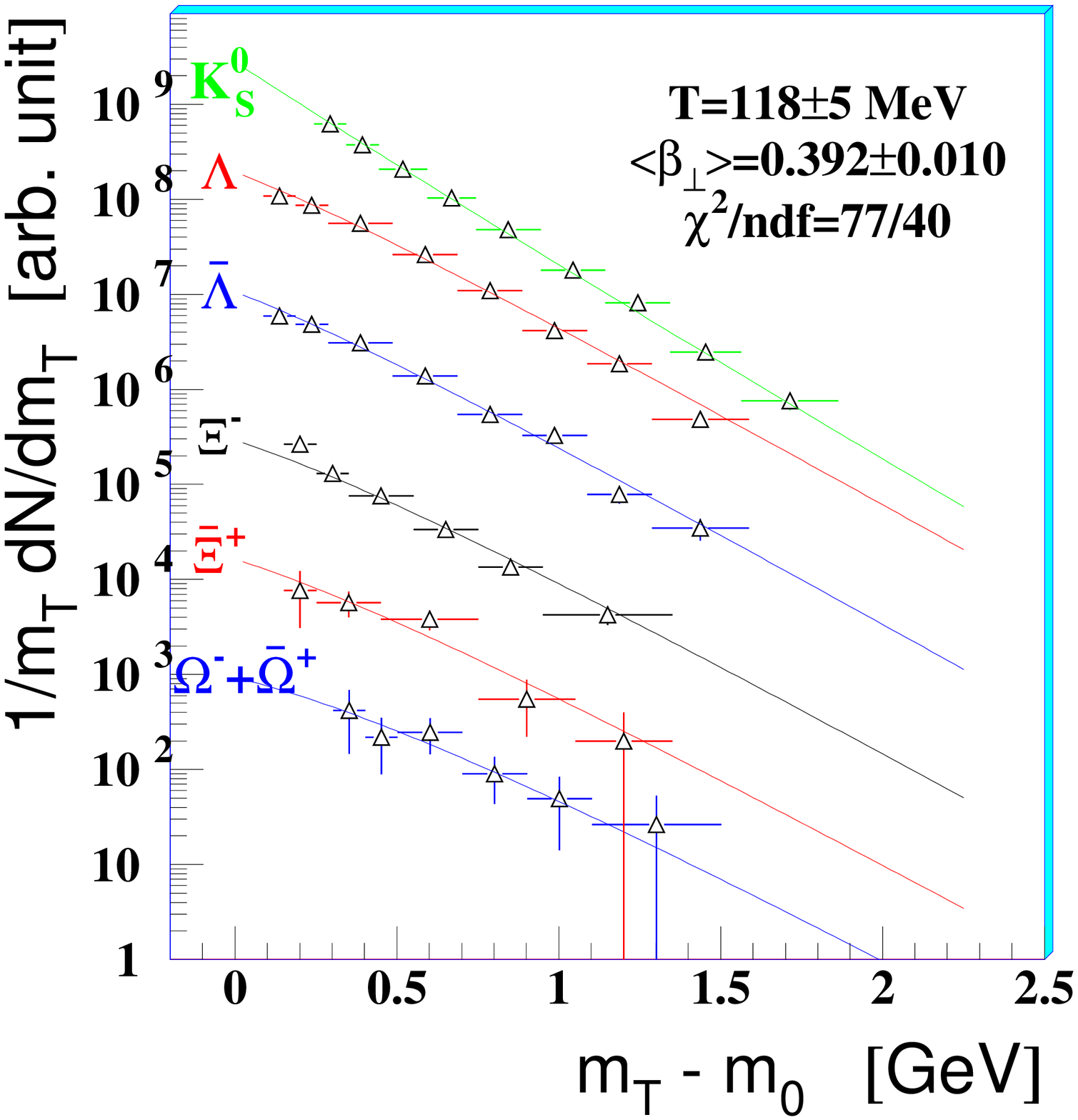}
\includegraphics{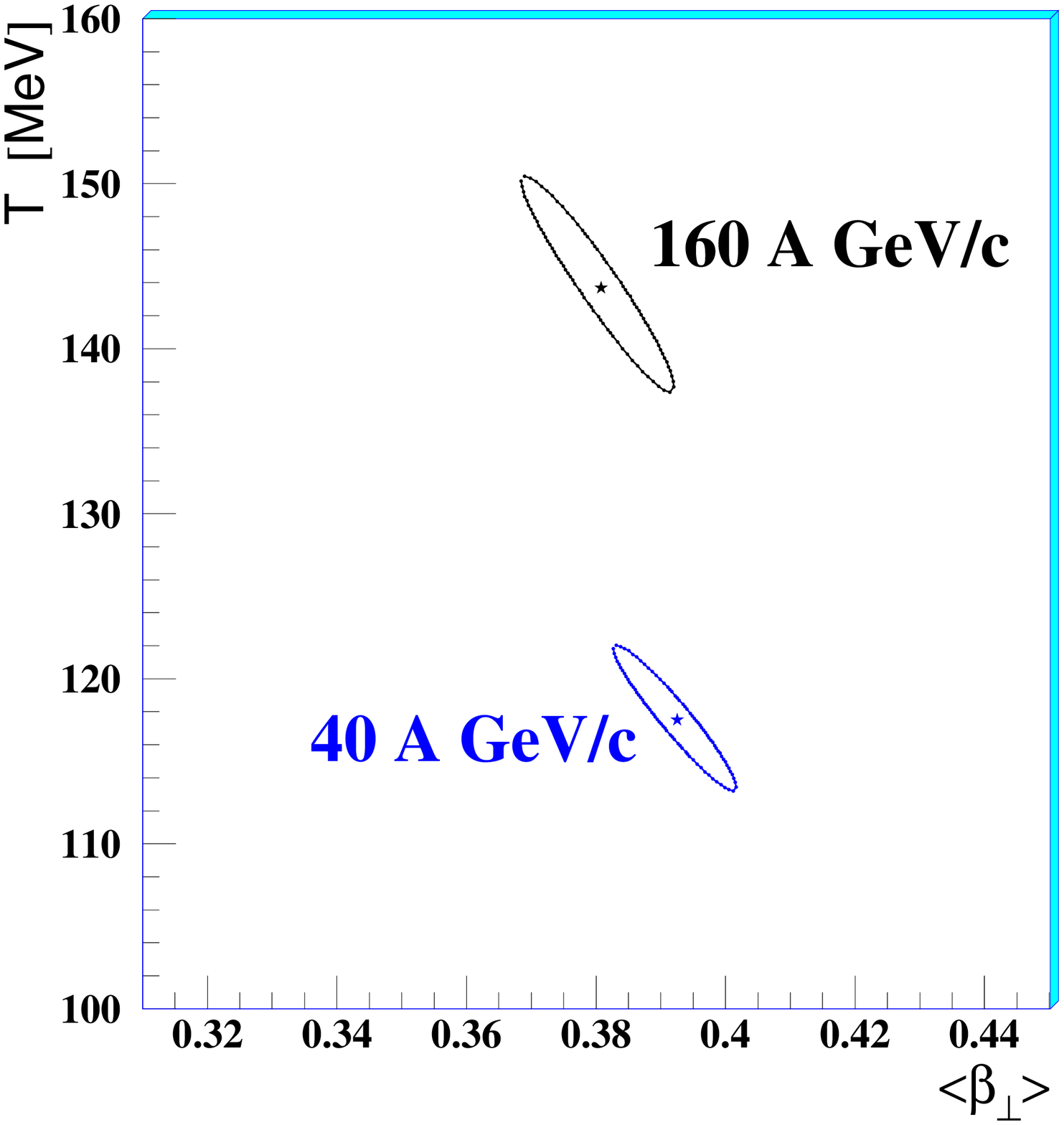}
}
\caption{ Left: blast-wave fits to the
          transverse mass spectra of strange particles for the 
          most central 53\% of the Pb--Pb inelastic cross-section at 40  $A$\ GeV/$c$.  \\
          Right: contour plots in the $T$--$\Bt$\  
          plane at the 1$\sigma$\  confidence level
          %evaluated at $\chi^2=\chi^2_{min}+1$\
          as compared to the result at 158 $A$\ GeV/$c$~\cite{BlastPaper} in
          the same centrality range.
          }
\label{fig:spettri_blast}
\end{figure}

The result of the global fit of equation~\ref{eq:Blast} with %$n=1$\ 
a linear profile hypothesis (i.e. with the exponent $n=1$\ in equation~\ref{eq:profile})  
to the data points of all measured strange  
particle spectra is shown in figure~\ref{fig:spettri_blast} (left panel) for the event 
sample corresponding to the most central 53\% of the inelastic Pb--Pb cross-section; the fit   
yields the following values for the two parameters $T$\ and $\Bt$:
\[ \fl
T = 118 \pm 5 {\tt (stat)} ^{+11}_{-10} {\tt (syst)} {\rm MeV} \, , \quad %\quad
% <\beta_\perp>=0.392 \pm 0.010 {\tt (stat)} ^{+0.011}_{-0.013} {\tt (syst)} 
 \Bt=0.392 \pm 0.010 {\tt (stat)} ^{+0.013}_{-0.015} {\tt (syst)} 
\nonumber
\]
%with $\chi^2/{\rm ndf} = 77/34$.   
with $\chi^2/{\rm ndf} = 77/40$.   
A large contribution to the $\chi^2$\ comes from  
the $\Xi$ spectra: the possibility of an  %scenario with an  
early freeze-out of multi-strange particles is discussed below.

The $T$\ and $\Bt$\ parameters are statistically
anti-correlated, as can be seen from the confidence level contour shown
in figure~\ref{fig:spettri_blast} (right panel). 
The systematic errors on $T$\  
and $\Bt$\  are instead correlated.  
%and they are estimated to be about $11\%$\ and $4\%$, respectively.

The results 
of the fits  
%of fits to equation~\ref{eq:profile}  
with different profile hypotheses  
%(i.e. different values of the exponent $n$\ in equation~\ref{eq:profile}) 
are given in table~\ref{tab:profile}.
\begin{table}[htb]
\caption{Results of the blast-wave model fit using different
         transverse velocity profiles. The quoted errors are statistical.
         The systematic errors on the temperature and on the velocity  
         are estimated to be about $11\%$\ and $4\%$, respectively,
         for all four profiles.
\label{tab:profile}}
\begin{center}
\begin{tabular}{lllll}
\hline
%                        updated         updated          updated           updated            
                    & {\bf $n=0$}     & {\bf $n=1/2$}   & {\bf $n=1$}     & {\bf $n=2$}   \\ \hline
%                          \multicolumn{5}{c}{\bf  40 $A$\ GeV/$c$ } \\
 $T$\ (MeV)         &  $125 \pm 4$    & $ 121\pm4 $     & $ 118\pm5 $     & $135 \pm 10$     \\
 $\Bt$  & $0.434\pm0.011$ & $0.422\pm0.011$ & $0.392\pm0.010$ & $0.302\pm0.016$ \\
% $\chi^2/{\rm ndf}$  &   $ 41/34 $     &    $ 51/34 $    & $ 77/34 $       &  $167/34 $      \\
 $\chi^2/{\rm ndf}$  &   $ 41/40  $     &    $ 51/40 $    & $ 77/40 $       &  $167/40 $      \\
%\hline
%                          \multicolumn{5}{c}{\bf 158 $A$\ GeV/$c$ } \\
% $T$\ (MeV)         &  $158 \pm 6$    & $ 152\pm6 $     & $ 144\pm7 $     & $151 \pm 11$    \\
% $<\beta_{\perp}>$  & $0.396\pm0.015$ & $0.394\pm0.013$ & $0.381\pm0.013$ & $0.316\pm0.014$ \\
% $\chi^2/ndf     $  & $ 39.6/48 $     &  $ 36.9/48 $    & $ 37.2/48 $     & $68.0/48 $      \\
\hline
\end{tabular}
\end{center}
\end{table}
\noindent
%The values of $T$\ and $<\beta_\perp>$\ are found to be statistically
%anti-correlated; the systematic errors on $T$\   
%and $<\beta_{\perp}>$, instead, are correlated: 
%they are estimated to be $10\%$\ and $3\%$, respectively. 
Contrary to the 158 $A$\ GeV/$c$\ case~\cite{BlastPaper}, where the use of the 
three profiles $n$=0, 1/2 and 1 results in equally well described 
spectra\footnote{The profile $n=2$\ was instead disfavoured 
by data.} with similar values of the freeze-out temperature  
and of the average transverse flow velocity, at this energy the quality 
of the fit is best with $n=0$\ and gradually 
%declines with increasing the exponent $n$.
worsens as the exponent $n$\ is increased.  
Since the case $n=0$\ is 
unphysical, the choice  $n=1/2$, which has also been suggested  
%to resemble closely 
to be a good approximation of the full hydro-dynamical  calculation~\cite{WiedPrivate}, 
appears to be favoured.  
However, for the sake of comparison with the higher energy results, %where $n=1$\ was used   
in the following a linear ($n=1$) radial dependence of the transverse flow velocity is used.  

The 1$\sigma$\ contour plot in the freeze-out parameter space is compared   
to the 158 $A$\ GeV/$c$\ result~\cite{BlastPaper} in the right-hand panel of 
figure~\ref{fig:spettri_blast}.  
%The global fits of equation~\ref{eq:Blast} %with $n=1$\  
%to the data points of all the measured strange particle spectra and the 
%corresponding 1$\sigma$\ contour plots  
%are shown in figure~\ref{fig:spettri_blast}.  
A lower thermal freeze-out temperature is measured at the lower beam energy, while   
the transverse flow velocities are found to be compatible within the errors.  
%
%\begin{figure}[hbt]
%\centering
%\resizebox{0.45\textwidth}{!}{%
%\includegraphics{cont_Energy.eps}}
%\caption{\rm Contour plots in the $<\beta_{\perp}>$--$T$\ plane at 1$\sigma$\
%             confidence level for the 53\% most central Pb--Pb collisions at
%             40 and 158 $A$\ GeV/$c$.}
%\label{fig:cont_ene}
%\end{figure}

\subsection{Particles with/without quarks in common with the nucleon.}
Since the particles which share valence quarks with the nucleons 
may have a different behaviour %than those 
%with respect to those %particles  
from  
which do not, we have performed  
separate fits for the particles of the two groups.  
Results of separate blast-wave fits are given 
in table~\ref{tab:Blast1}. 
\begin{table}[h]
\caption{Thermal freeze-out temperature  
and average transverse flow velocity in the full centrality range.    
%assuming a linear transverse profile ($n=1$).
The first error is statistical, the second one systematic.
\label{tab:Blast1}}
\begin{center}
\begin{tabular}{lccc}
\hline
particles & $T$ (MeV) &  $\Bt $ & $\chi^2/{\rm ndf}$ \\ \hline\hline
{\bf $K_S^0$}, {\bf $\La$},  {\bf $\Xi^-$} &
%$ 120 \pm 7 \pm 12 $ & $ 0.39 \pm 0.02\pm 0.01  $ & $ 56.2/18 $ \\ 
$ 120 \pm 7 \pm 13 $ & $ 0.39 \pm 0.02\pm 0.02  $ & $ 56.2/21 $ \\ 
{\bf $\Al$}, {\bf $\overline\Xi^+$}, {\bf $\Omega^-$}, {\bf $\overline\Omega^+$} &
%$ 80 \pm 22 \pm 11 $ & $ 0.45 \pm 0.03  \pm 0.01  $ & $ 14.8/18 $ \\
$ 80 \pm 22 \pm 11 $ & $ 0.45 \pm 0.03  \pm 0.02  $ & $ 14.8/17 $ \\
\hline
%   \multicolumn{4}{c}{\bf 158 $A$\ GeV/$c$ } \\
%{\bf $K_S^0$}, {\bf $\La$},  {\bf $\Xi^-$} &
%$146 \pm 8 \pm 14 $ & $ 0.376 \pm 0.015 \pm 0.012 $ & $ 18.1/23 $ \\
%{\bf $\Al$}, {\bf $\overline\Xi^+$}, {\bf $\Omega^-$}, {\bf $\overline\Omega^+$} &
%$ 130\pm28 \pm 14 $ & $ 0.403 \pm 0.032 \pm 0.012 $ & $ 18.5/23 $ \\ \hline
\end{tabular}
\end{center}
\end{table}
\noindent
The freeze-out conditions for the two groups of particles are compatible within 2$\sigma$.   
Since the interaction cross-sections for
the particles of the two groups are expected to be very different, this finding would suggest  
%again
limited importance of final state interactions  
(i.e. a rapid thermal freeze-out) and  
 similar production mechanisms for the two groups. 
A similar conclusion has also been drawn at 158~$A$ GeV/$c$~\cite{BlastPaper}, where   
the freeze-out parameters for the two groups agree within 1$\sigma$.  
\subsection{Earlier freeze-out of multi-strange particles ?}  
At higher energies it has been argued that multi-strange baryons may undergo 
an earlier freeze-out than other particles: at top SPS energy this scenario  
has been suggested based on the $\Omega$\ spectra measurements of  
the WA97~\cite{MtWA97}, NA57~\cite{BlastPaper} and NA49~\cite{NA49Omega} Collaborations.  
At RHIC, results from the STAR Collaboration~\cite{STARstrange} suggest that the  
$\Xi$\ particles freezes out earlier than  $\pi$, K, p and $\Lambda$.  

\begin{figure}[b]
\centering
%\resizebox{0.90\textwidth}{!}{%
%\includegraphics{cont_Strange2.eps}
%\includegraphics{InvSlop_mass_fin.eps}}\\
\resizebox{0.90\textwidth}{!}{%
\includegraphics{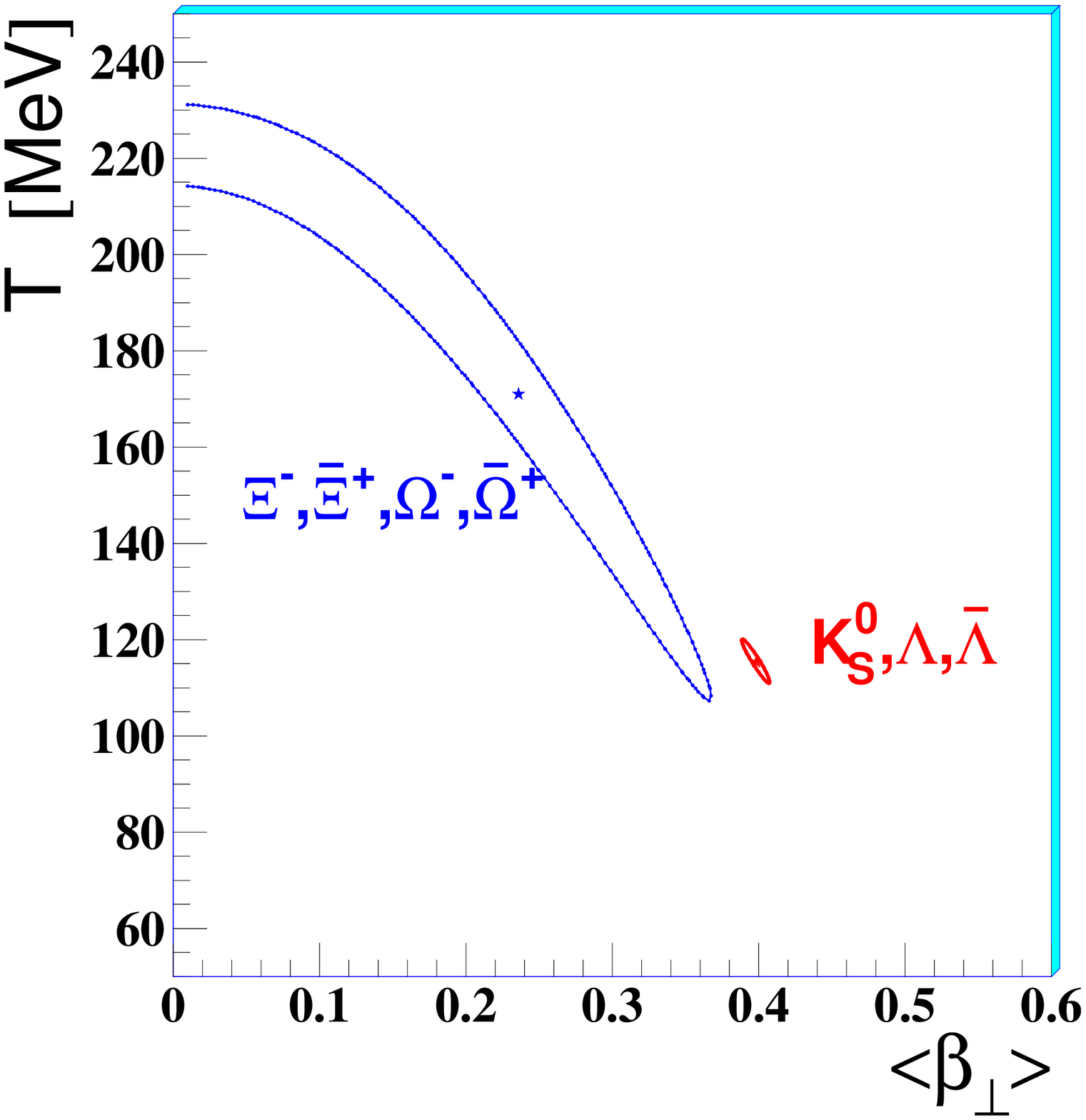}
\includegraphics{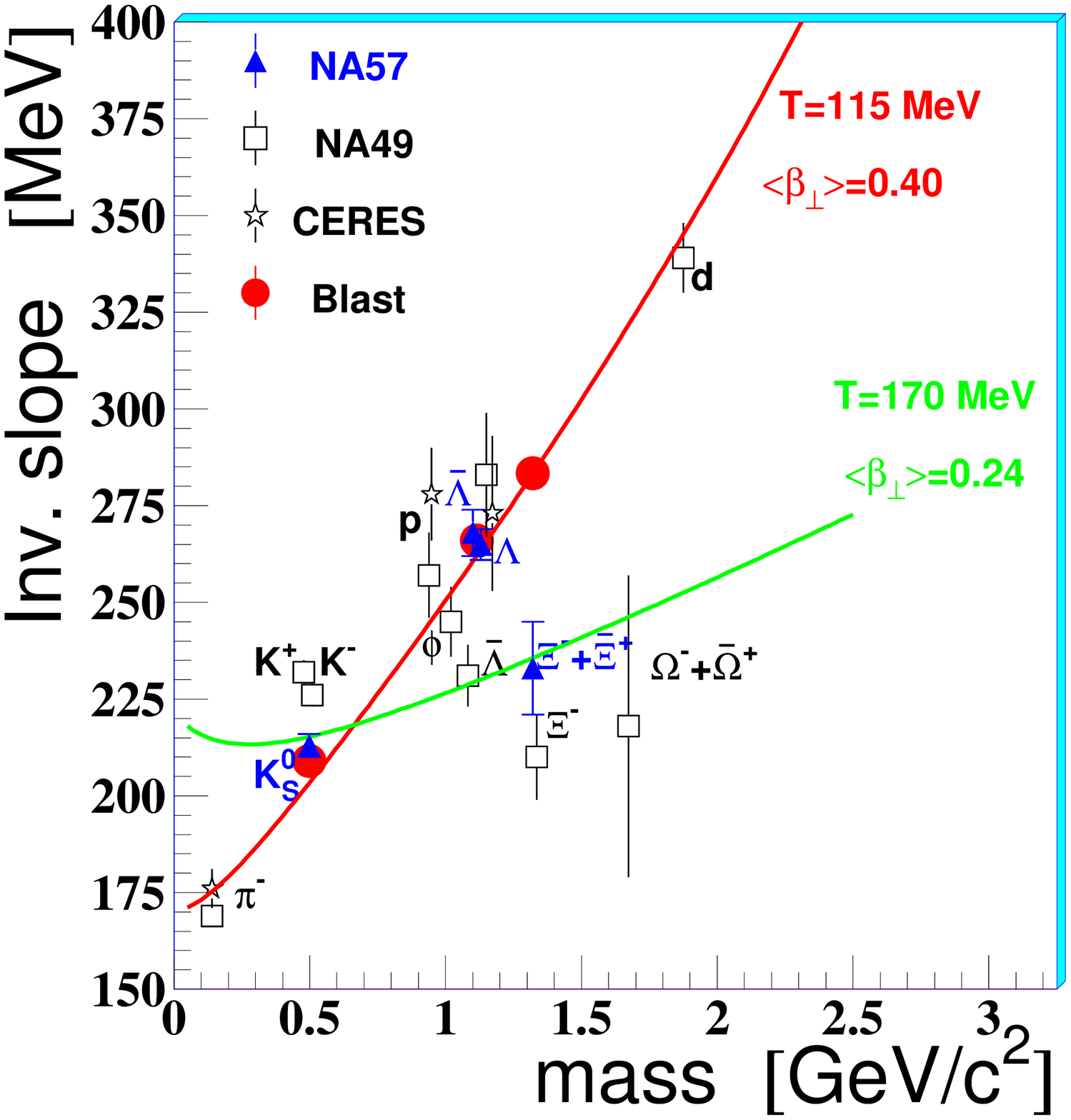}}
\caption{%Top: 158 $A$\ GeV/$c$,  bottom: 40 $A$\ GeV/$c$.\\
         Left: the thermal freeze-out temperature versus the average transverse flow
         velocity for blast-wave fits using a linear ($n=1$) velocity profile.
         The 1$\sigma$\ contours are shown, with the markers indicating
         the optimal fit locations.
         Right: prediction of the blast-wave model for
         inverse slopes (see text for details).
         % For data references see~\cite{BlastPaper}.
         }
\label{fig:BlastPred}
\end{figure}

The 1$\sigma$\ contours of the separate blast-wave fits for singly  
and multiply-strange particles are shown in the left-hand panel 
of figure~\ref{fig:BlastPred}. 
%
%\begin{figure}[hbt]
%\centering
%\resizebox{0.90\textwidth}{!}{%
%\includegraphics{cont_strange.eps}
%\includegraphics{InvSlopVsMass_Blast_cont_b1.eps}}
%\caption{Left: the thermal freeze-out temperature vs the average transverse flow
%         velocity for blast-wave fits using a linear ($n=1$) velocity profile.
%         The 1$\sigma$\ contours are shown for singly-strange and multi-strange
%         particles, with the markers indicating he optimal fit locations.
%         Right: prediction of the blast-wave model for
%         inverse slopes at 40 $A$\ GeV/$c$\ (see text for details).}
%\label{fig:BlastPred_40}
%\end{figure}
%
%The results of the two data sets do not overlap thus indicating that the $\PgXm$\ 
The plot seems to indicate that the $\PgXm$\ 
hyperon, which statistically dominates the multiply-strange particle fit, 
%may show indeed  
may have  
a different thermal freeze-out behaviour from \PKzS, \PgL\ and \PagL. 
However,  
%
%groups of particles are compatible with the result of the global fit  
%determination.
%However, the fit for the multiply strange particles is statistically
%dominated by the $\Xi$;  in fact the $\Xi+\Omega$\
%contour remains essentially unchanged when fitting the $\Xi$\ alone.
%Therefore we can only conclude
%that the $\Xi$\ undergoes a thermal freeze-out which is compatible with
%that of $\Lambda$\ and $\PKzS$.
%For the $\Omega$, 
due to the low statistics,   
it is not possible to extract significant values
for both freeze-out parameters  
from the \PgXm\ spectrum alone.    
The possibility of a deviation for the \PgXm\ hyperon 
from the freeze-out systematics extracted from the combined fit to the 
$\PKzS$, $\Lambda$\ and \PagL\  spectra
can be better inferred from the integrated information of the
\PgXm\ spectrum, i.e. from its inverse slope. 
In figure~\ref{fig:BlastPred} (right panel) we plot a compilation of 
data\footnote{
The NA49 results are taken from the following references: 
\Pgpm, \PKp\ and \PKm\ from~\cite{NA49pions}; 
\Pp\ and deuteron from~\cite{NA49ProtDeut}; 
\PgL\ and \PagL\ from~\cite{NA49Lambda}; 
$\phi$\ from~\cite{NA49Phi}; \PgXm\ from~\cite{NA49Xi} and $\Omega$\ from~\cite{NA49Omega}.   
The CERES inverse slopes, which are taken from reference~\cite{Ceres}, 
are shown for negative hadrons, net proton and \PgL.} 
on the $m_{\tt T}$\ inverse slopes 
measured in Pb--Pb collisions at 40 $A$\ GeV/$c$,  superimposed to the blast-wave 
model results.  
The full lines represent the inverse slope one would obtain by fitting  
an exponential to a ``blast--like'' $1/m_{\tt T} \, dN/dm_{\tt T}$\ 
distribution (i.e. to equation~\ref{eq:Blast}) for a generic particle of mass $ m_{0} $,    
in the common range $ 0.05 < m_{\tt T} - m_{0}  < 1.50$\ GeV/$c^2$,   
for two different freeze-out conditions.   
%absence of transverse flow ($<\beta_{\perp}>=0$) and our best fit determination.  
The first corresponds to the parameters of the best fit to the singly-strange 
particles ($T=115$\ MeV , $\Bt=0.40$), the second to those of the 
multiply-strange ones ($T=170$\ MeV, $\Bt=0.24$).   
Since the inverse slope is a function of the $m_{\tt T} - m_{0}$\ 
range where the fit is performed, we have also computed the blast-wave 
inverse slopes of $\PKzS$, $\PgL$\ and $\Xi$\ %and $\Omega$\  
spectra in the $m_{\tt T} - m_{0}$\ ranges of NA57  
(closed circles).  
The measured values of the inverse slope of the multi-strange baryons appear 
to deviate significantly from the trend of the other strange particles.  

%At 40 $A$\ GeV/$c$\ the same analysis suggests an early decoupling even  
%of the $\Xi$\ with respect to the singly strange particles  
%(see figure~\ref{fig:BlastPred}). 
%
\subsection{Centrality dependence}
The hydro-dynamical description 
%(or the description of model inspired to hydro-dynamics) 
of observables related to collective dynamics, e.g.  
the elliptic flow, is strongly influenced by the freeze-out temperature.  
It is therefore important to determine how the thermal freeze-out conditions may 
change with respect to the initial collision geometry.  

%collision centrality.
The centrality dependence of the freeze-out parameters at 158 $A$\ GeV/$c$\ 
beam momentum can be summarized as follows:   
the more central the collisions the larger the transverse collective  
flow and the lower the final thermal freeze-out  
temperature~\cite{BlastPaper}, the dependence being more pronounced for 
the flow.  
A similar behaviour was reported at RHIC  by the PHENIX~\cite{PHENIXcen}, 
STAR~\cite{STARstrange} and BRAHMS~\cite{BRHAMScen} Collaborations.    
Higher freeze-out temperatures for more peripheral collisions may be interpreted
as the result of earlier decoupling of the expanding system.  

%Therefore, when trying to describe hydro-dynamically  
%the experimental data measured for peripheral or semi-central collisions, 
%one should employ higher values of the freeze-out temperature than those 
%measured for central collisions.   
%Indeed, in reference~\cite{NA45} the measured  
%elliptic flow for Pb--Pb at 158 $A$\ GeV/$c$\ in the centrality range 
%$\sigma/\sigma_{inel}=(13-26)\%$\ disagrees to that obtained with a hydro-dynamical 
%evolution terminated at $T=120$\ MeV 
%(%measured 
%a well established 
%freeze-out temperature for {\em central} collisions at top SPS energy, see 
%reference~\cite{BlastPaper} for a compilation of results)  
%but it is close to that terminated at $T=160$\ MeV\footnote{
%In the range $\sigma/\sigma_{inel}=(11-23)\%$\ the NA57 measurements gives 
%$T=146\pm17$\ MeV~\cite{BlastPaper}.}.
%

\begin{figure}[t]
\centering
\resizebox{0.45\textwidth}{!}{%
\includegraphics{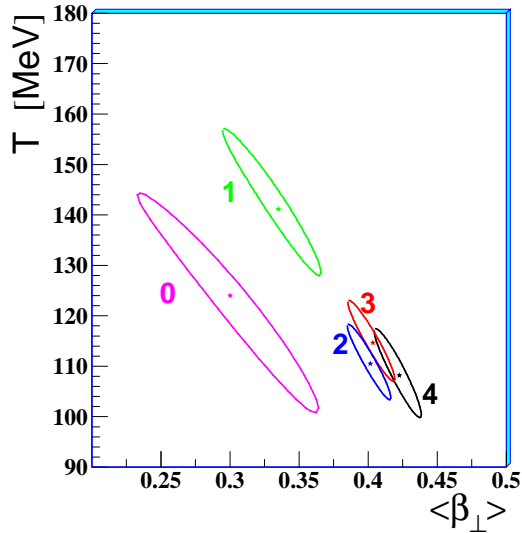}}
\caption{%The~1$\sigma$~contours~%from fits in each centrality class
         The 1$\sigma$ confidence level contours from fits
         in each centrality class.}
\label{fig:cont_msd}
\end{figure}

In figure~\ref{fig:cont_msd} we show the $1\sigma$\ confidence level contours
for each of the five centrality classes defined in table~\ref{tab:centrality}.
%in the $<\beta_\perp>$--$T$\ plane
%as obtained for the $n=1$\  profile.
%Also at 40 $A$\ GeV/$c$\ beam momentum,  
With decreasing collision centrality the transverse flow velocity decreases 
steadily, as also observed at higher energies. The freeze-out temperature is roughly 
constant within the errors at the value $T \approx 110 \pm 10$\ MeV for the three 
most central classes  (i.e. for the most central 23\% of the inleastic Pb--Pb 
cross-section). %and then it increases to the value $T=141\pm14$\ MeV for class 1.  

\section{Conclusions}
We have analyzed the transverse mass spectra of \PgL, \PgXm, \PgOm\ hyperons,  
their anti-particles and \PKzS\ mesons in Pb--Pb  collisions at 40 $A$\ GeV/$c$\ beam momentum over a centrality 
range corresponding to the most central 53\% of the Pb--Pb inelastic cross-section.  

The inverse slopes of these particles 
are found to be lower than those measured at 158 $A$\ GeV/$c$\ beam momentum.  
%by about 
As a function of centrality, the inverse slopes of the singly-strange particles 
%show a %significant 
%step-like  
%increase when going from the most peripheral to the other centrality classes; 
increase significantly when going from the most peripheral class to the next one;  
%and then they remain constant; 
for the more central classes, a weak increase with centrality cannot be excluded for \PgL\ and \PagL.   
The inverse slope of the \PgXm\ is compatible with being constant over the whole centrality range. 
\PagXp\ and $\Omega$\ hyperons have not been studied as a function of the centrality due to 
the limited statistics.  

Particle and anti-particle inverse slopes are compatible within the errors; 
in particular, the inverse slopes of \PgL\ and \PagL\ hyperons 
%are compatible within small relative errors 
%are, to  a great precision (5\%), the same  
are the same within 5\%  
over the  %explored 
covered    
centrality 
range, thus suggesting that strange baryons and anti-baryons may be produced and  
evolve in the collision dynamics by similar mechanisms.  

The inverse slope of the \PgXm\ hyperons significantly deviates from the
general trend of values increasing with the particle rest mass, as 
observed for non-strange and singly-strange particles.

The analysis of the transverse mass spectra %at 158 $A$\ GeV/$c$\ 
%of strange particles in Pb--Pb collisions at SPS energies  
in the framework of the blast-wave  model 
suggests  that after a central 
collision the system expands   
with an average transverse  flow
velocity of about 40\% of the speed of light
and then it freezes out when the temperature is of the order of 110 MeV.   
The measured transverse flow velocity  is compatible with that measured at 158 $A$\ GeV/$c$\ 
but the freeze-out temperature is lower at  low energy. 
The inverse %slopes of multi-strange particles measured by NA49 and NA57 deviate from the values 
slope of the \PgXm\ hyperon deviates from the value 
%($\Omega$\ at 158 $A$\ GeV/$c$, $\Xi$\ and $\Omega$\ at 40 $A$\ GeV/$c$)   
%the $\Omega$\ particle 
%appear to 
predicted by the blast-wave model tuned on singly-strange particles 
(\PKzS, \PgL\ and \PagL).    
Finally, the  results on the centrality dependence of the  
expansion dynamics  
indicate that with increasing centrality  
the transverse flow velocity increases steadily 
and the freeze-out temperature decreases when going from class 1 (23--40\%) 
to the more central classes.  
%
%%%%%%%%%%%%%%%%%%%%%%%%%%%%%%%%%%%%%%%%%%%%%%%%%%%%
%
%%%%%%%%%%%%%%%%%%%%%%%%%%%%%%%%%%%%%%%%%%%%%%%%%%%%%%%%%%
%\ack
%We are grateful to U~Heinz, J~Rafelski, B~Tom\'{a}\v{s}ik and U~A~Wiedemann
%for useful comments and fruitful discussions.   
%%%%%%%%%%%%%%%%%%%%%%%%%%%%%%%%%%%%%%%%%%%%%%%%%%%%%%%%%%%%%%%%%%%%%%%%%
%%%%%%%%%%%%%%%%%%%%%%%%  BIBLIOGRAFIA %%%%%%%%%%%%%%%%%%%%%%%%%%%%%%%%%%

%%%

\section*{References}
\begin{thebibliography}{33}
%
%%%%%%%%%%%%%%%%%%%%%%%%%% Intro %%%%%%%%%%%%%%%%%%%%%%%%%%
%
\bibitem{lattice} Karsch F 2002 {\it Lect. Notes Phys.} {\bf 583} 209 
\bibitem{QM04-QM05} Ritter H G and Wang X-N (ed) 2004 \JPG {\bf 30} S633-S1430 
({\it Proc. Quark Matter 2004}) \nonum
%\bibitem{QM05}  
L\'{e}vai P and Cs\"{o}rgo T (ed) 2006 \NP A  in press  
({\it Proc. Quark Matter 2005}) 
\bibitem{rafSQM03} Rafelski 2004 \JPG {\bf 30} S1-S28
\bibitem{rafelski} Rafelski J and M\"{u}ller B 1982 \PRL {\bf 48} 1066 \nonum
                   Rafelski J and M\"{u}ller B 1986 \PRL {\bf 56} 2334 \nonum
                   Koch P, M\"{u}ller B and Rafelski J 1986 {\it Phys. Rep.} {\bf 142} 167 \nonum
                   Rafelski J 1991 \PL B {\bf 262} 333
\bibitem{enh160} Antinori F {\it et al.} 2006 \JPG {\bf 32} 427-441
\bibitem{WA97enh} Andersen E {\it et al.} 1999 \PL B {\bf 449} 401 \nonum
                  Antinori F \etal 1999 \NP A {\bf 661} 130c
\bibitem{NA57QM04} Bruno G E {\it et al.}  2004 \JPG {\bf 30} S717-S724
\bibitem{NA49pions}  Afanasiev S V {\it et al.} 2002 \PR C {\bf 66} 054902 %kaon/pi
\bibitem{NA49Lambda} Antic T \etal 2004 \PRL {\bf 93} 022302 %lambda/pi
\bibitem{NA49str} Alt C \etal 2005 \PRL {\bf 94} 052301 %k,lambda,phi in C-C and Si-Si
%\bibitem{NA49Xi} Afanasiev S V {\it et al.}  2002 \PL B {\bf 538} 275-281 %Xi/pi
%\bibitem{NA49Omega} Alt C  {\it et al.} 2005 \PRL {\bf 94} 192301 %Omega/pi
\bibitem{ThermalMod} Andronic A, Braun-Munzinger P and Stachel J 2006 \NP A {\bf 772} in press, nucl-th/0511071 \nonum
                     Becattini F, M. Ga\'zdzicki M, Keranen A, Manninen J and Stock R 2004 \PR C {\bf 69} 024905 \nonum
                     Becattini F, Manninen J and Ga\'zdzicki 2006 \PR C {\bf 73} 044905 \nonum
                     Letessier J and Rafelski J 2005 submitted to \PR C, nucl-th/0504028 \nonum 
	             Broniowski W, Florkowski W and Michalec M 2002 {\it Acta Phys. Pol.} B {\bf 33} 761, 
                            nucl-th/0106009 \nonum
                     Broniowski W and  Florkowski W 2002 \PR C {\bf 65} 064905, nucl-th/0112043
\bibitem{ReviewHydro} 
   Hirano T 2004 \JPG {\bf 30} S845-S851 \nonum
   Torrieri G and Rafelski J \JPG {\bf 30} S557-S564 \nonum
   Heinz U W 2005 \JPG {\bf 31} S717-S724 \nonum
   Csernai L P, Moln\'ar E, Ny\'iri \'A and Tamosiunas K 2005 \JPG {\bf 31} S951-S957 \nonum
   Steinberg P A 2005  \NP A {\bf 752} 423c-432c
\bibitem{BlastRef} Schnedermann E, Sollfrank J and Heinz U 1993 \PR C
                   {\bf 48} 2462 \nonum
                   Schnedermann E, Sollfrank J and Heinz U 1994 \PR C {\bf 50} 1675
\bibitem{BlastPaper} Antinori F {\it et al.} 2004 \JPG {\bf 30} 823-840 
\bibitem{MANZ} Manzari V {\it et al.} 1999 \JPG {\bf 25} 473   \nonum
               Manzari V {\it et al.} 1999 \NP A {\bf 661} 761c
%\bibitem{Omega2}  Campbell M \etal 1994 \NIM  A {\bf 342} 52
%\bibitem{Omega3}  Heijne E H M \etal 1996 \NIM A {\bf 383} 55
\bibitem{Multiplicity} Antinori F {\it et al.} 2005 \JPG {\bf 31} 321-335
%
\bibitem{PDG} Eidelman S \etal  2004 {\it Review of Particle Properties} \PL B {\bf 592} 1
\bibitem{BrunoMoriond02} Bruno G E {\it et al.} 2002 in {\it QCD and High Energy Hadronic Interactions} 
             edited by Tr\^{a}n Thanh V\^{a}n 405-410 ({\it Proc. $37^{th}$\ Rencontres de Moriond}); 
             {\it Preprint} hep-ex/0207047 
\bibitem{RapPaper} Antinori F {\it et al.}  2005 \JPG {\bf} 31 1345-1357
% \nonum      Bruno G E 2002 {\it Ph.D. Thesis}, University of Bari
%            ({\it Preprint} nucl-ex/0402014)
%\bibitem{WoundNucl} Bialas A, Bleazy\'nski M and Czy\.{z} W 1976 \NP B {\bf 111} 461   
%		   see also www.cern.ch/CERN/Announcements/2000/NewStateMatter
%
\bibitem{SecondRef} Lee K S, Heinz U and Schnedermann E 1990 \ZP C {\bf 48} 525 \nonum 
                    Cs\"{o}rgo T and Lorstad B 1996 \PR C {\bf 54} 1390  
\bibitem{MtWA97} Antinori F {\it et al.} 2000 {\it Eur. Phys. J.} C {\bf 14} 633
\bibitem{NA49Omega} Alt C  {\it et al.} 2005 \PRL {\bf 94} 192301
\bibitem{Hecke} van~Hecke~H, Sorge~H and Xu~N 1998 \PRL {\bf 81} 5764
%
%
%\bibitem{NA49pions}  Afanasiev S V {\it et al.} 2002 \PR C {\bf 66} 054902 
\bibitem{NA49Xi} Afanasiev S V {\it et al.}  2002 \PL B {\bf 538} 275-281
%\bibitem{NA49Omega} Alt C  {\it et al.} 2005 \PRL {\bf 94} 192301 
\bibitem{NA49Xi40} Meurer C {\it et al.} 2004 \JPG {\bf 30} S1325-S1328
%
\bibitem{35} Cooper F and Frye G 1974 \PR D {\bf 10} 186
%\bibitem{RapPaper} Antinori F {\it et al.}  2005 \JPG {\bf} 31 1345-1357
\bibitem{WiedPrivate} Wiedemann U A 2003 {\it Private communication}
\bibitem{STARstrange}  Adams J. {\it et al.} 2004 \PRL {\bf 92} 182301
%
\bibitem{NA49ProtDeut} Anticic T {\it et al.} 2004 \PR C {\bf 69} 024902  
%\bibitem{NA49Lambda}   Anticic T {\it et al.} 2004 \PRL {\bf 93} 022302
\bibitem{NA49Phi}  Friese V {\it et al.} 2005 \JPG {\bf 31} S911-S918 
and {\it private communication}   
\bibitem{Ceres} Schmitz W {\it et al.} 2002 \JPG {\bf 28} 1861-1868     
%
\bibitem{PHENIXcen} Adcox K {\it et al} 2004 \PR C {\bf 69} 024904
%\bibitem{PHOBOScen} 
\bibitem{BRHAMScen} Arsene I {\it et al} 2005 \PR C {\bf 72} 014908 
%
%\bibitem{NA45} Agakichiev G. {\it et al.} 2004 \PRL {\bf 92} 032301
%
%\bibitem{NA57proposal} Caliandro R {\it et al.}, NA57 proposal, 1996 
%{\it CERN/SPSLC 96-40, SPSLC/P300} 
%\bibitem{EliaHQ04} Elia D {\it et al.}, contribution to these proceedings
%
%
%\bibitem{Slope-p} Fini R A {\it et al.} 2001 \NP A {\bf 681} 141c
%\bibitem{Torrieri} Torrieri G and Rafelski J 2004 \JPG {\bf 30}  s557  \nonum 
%		   Torrieri G and Rafelski J 2002 {\it Preprint} nucl-th/0212091
%
%
%%%%%%%%%%%%%%%% Blast wave description of the spectra
%
%\bibitem{Nonso}   Esumi S, Chapman S, van Hecke H and Xu N 1997 {\PR} C {\bf 55} 2163
%%%%%%%%%%%%%%
\end{thebibliography}
\end{document}